\documentclass[prx, reprint, superscriptaddress, floatfix, nofootinbib]{revtex4-2}
\usepackage{graphicx}
\usepackage{subfigure}
\usepackage{nicefrac}
\usepackage{amsmath}
\usepackage{amssymb}
\usepackage{amsfonts}
\usepackage{amsthm}
\usepackage{mathrsfs}
\usepackage{wasysym}
\usepackage{bm}
\usepackage{placeins}
\usepackage[bookmarks=false]{hyperref}
\usepackage[usenames,dvipsnames]{xcolor}
\usepackage[italicdiff]{physics}
\usepackage{chemformula}
\usepackage{mathtools}
\usepackage{enumitem}
\usepackage{stackengine}
\usepackage[dvipsnames]{xcolor}
\usepackage{soul}

\usepackage{amsmath}

\DeclareMathOperator{\sgn}{sgn}

\newcommand{\hc}{\text{H.c.}}
\newcommand{\dg}{\dagger}
\newcommand{\up}{\uparrow}
\newcommand{\down}{\downarrow}

\newtheorem*{thm*}{Theorem}

\makeatletter
\newcommand{\subalign}[1]{%
  \vcenter{%
    \Let@ \restore@math@cr \default@tag
    \baselineskip\fontdimen10 \scriptfont\tw@
    \advance\baselineskip\fontdimen12 \scriptfont\tw@
    \lineskip\thr@@\fontdimen8 \scriptfont\thr@@
    \lineskiplimit\lineskip
    \ialign{\hfil$\m@th\scriptstyle##$&$\m@th\scriptstyle{}##$\hfil\crcr
      #1\crcr
    }%
  }%
}
\makeatother

\begin{document}

\title{
How quantum phases on cylinders approach the 2d limit}
\author{Yuval Gannot}
\affiliation{Department of Physics, Stanford University, Stanford, California 94305, USA}
\author{Steven A. Kivelson}
\affiliation{Department of Physics, Stanford University, Stanford, California 94305, USA}
\date{\today}

\begin{abstract}

We consider the properties of $T=0$ quantum phases of matter -- especially superconducting and analogous spin-liquid phases -- on infinite cylinders of width $L_\perp$ and analyze the ways in which the $L_\perp \to \infty$ (2d) limit is approached.  This problem is interesting in its own right, but is particularly important in the context of extrapolating accessible density matrix renormalization group (DMRG) results on model strongly interacting problems to the desired 2d limit.  Various methods for drawing firm conclusions about the quantum phases in 2d from relatively small $L_{\perp}$ results are  illustrated.
\end{abstract}

\maketitle

\section{Introduction}
\begin{figure}
\centering
\includegraphics[width=1.0\linewidth]{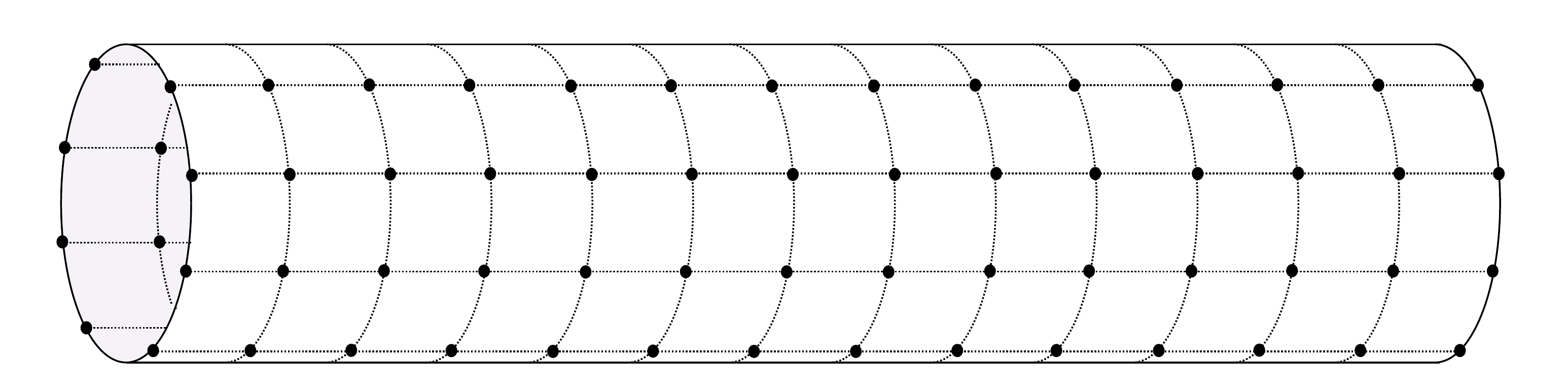}
\caption{Illustration of a segment of a cylinder for the case of a square lattice.
\label{fig:illustration}}
\end{figure}

Density matrix renormalization group (DMRG) is an efficient algorithm for computing the ground state properties of interacting quantum many-body systems in one spatial dimension (1d) \cite{White_1993,Schollwock_2011,McCulloch_2008}. 
For some time now, DMRG has successfully been used to study 1d ladders and cylinders of circumference ${L_{\perp}} > 1$ (measured in lattice sites; see also Fig.~\ref{fig:illustration} for an illustration.) \cite{Noack1994,Noack_White_Scalapino_1996,Stoudenmire_White_2012,Liu2012,Dodaro2017, Huang2018, Jiang2018, Jiang2018tj,
Jiang2019, Gong_Zhu_Sheng_2021, Jiang_Scalapino_White_2021,
Jiang_Kivelson_2021,Jiang_Yao_Balents_2012, Gong_et_al_2013, Gong_et_al_2014, Yan_Huse_White_2011, Jiang_Wang_Balents_2012,
Depenbrock_McCulloch_Schollwock_2012,
Kolley_et_al_2013,
Kolley_et_al_2015,
He_et_al_2017, 
Saadatmand_McCulloch_2016,
Saadatmand_McCulloch_2017,
Hu_et_al_2019, Jiang_Jiang_2022, Shirakawa_et_al_2017, Venderley_Kim_2019, Szasz_et_al_2020, Zheng_et_al_2020,Szasz_Motruk_2021,Peng_et_al_2021, Chatterjee_2020}.
As computational power has increased, so too has the largest accessible ${L_{\perp}}$. 
Consequently, there has  been an increasingly serious effort to extrapolate the results of these studies 
to ${L_{\perp}} = \infty$ 
to infer the properties of 2d systems.
This 
is challenging, however, since the presently accessible circumferences are still not very large. 

To facilitate this extrapolation, it is useful to have at hand a ``finite size scaling'' theory for how various candidate 2d phases behave when placed on cylinders of finite circumference. 
This was analyzed for the case of a Heisenberg quantum antiferromagnet by Chakravarty \cite{Chakravarty_1996}.  
He showed that if the ground state in the 2d limit is an antiferromagnet, then for cylinders of finite circumference,
the spin correlation length for even ${L_{\perp}}$ must diverge exponentially 
 as $\xi_{af} \sim \exp[\alpha {L_{\perp}}]$.
 \footnote{ Here $\alpha=2\pi \rho_s/c$ where $\rho_s\approx JS^2$ and $c\approx 2\sqrt{2}JS $ are, respectively, the spin-wave stiffness and velocity of the 2d AF, $J$ is the exchange coupling, and the approximate expressions are derived from leading order spin-wave theory  with $S$  the spin on each site.
Cylinders with
${L_{\perp}}$ odd and $S$ half-odd-integer are gapless due to a topological term, in agreement with the generalized Lieb-Schultz-Mattis theorem \cite{Lieb_Schultz_Mattis_1961,Yamanaka_Oshikawa_Affleck_1997}.} Thus, antiferromagnetic order in the 2d limit leaves an unambiguous imprint on cylinders of even moderate circumference. In the same spirit as Chakravarty's analysis, finite-$L_{\perp}$ results for various kinds of topological order were obtained in Refs.~\cite{Yao_Kivelson_2012,Jiang_Yao_Balents_2012,Seidel_et_al_2005,Strinati_et_al_2019}.

Here, we present an analogous finite size scaling theory for $s$-wave and $d$-wave superconductors. The general logic in this paper mirrors closely that of Chakravarty and Refs.~\cite{Yao_Kivelson_2012,Jiang_Yao_Balents_2012,Seidel_et_al_2005,Strinati_et_al_2019}: We start with the assumption that the 2d limit is a  superconductor, and derive from that assumption the scaling with ${L_{\perp}}$ of various observables.
We address the long-distance asymptotic behavior of the superconducting correlations and their dual, charge density wave (CDW) correlations, as well as the quasiparticle spectrum and various more microscopic properties.

We find a number of practically useful results: 1) Any system which is superconducting in the 2d limit should exhibit strong, power-law superconducting correlations with an exponent that rapidly approaches zero at large ${L_{\perp}}$;  superconductivity should not be hard to detect!
2) If the 2d superconductor has a nodeless gap,  introducing a small amount of anisotropy to the underlying band structure can substantially speed up the rate at which the two-dimensional limit is approached. 
3) If the 2d superconductor has a nodal gap,  this can readily be identified on a cylinder by exploiting twisted boundary conditions; specifically, the quasiparticle gap minimized over all possible twisted boundary conditions either vanishes identically, or is exponentially small in 
${L_{\perp}}$.
Finally, as is the case for other physical systems \cite{Lin_Zong_Ceperly_2001}, twisted boundary conditions can be exploited to speed up other aspects of the approach to the thermodynamic limit.

While our results are explicitly applicable only to superconducting (and CDW) orders, we conjecture that the results may be more generally useful.  Specifically, there are useful formal analogies between the theory of superconductors and various other phases of matter -- especially spin liquids. (For a review, see \cite{Broholm_et_al_2020}.)
It is 
plausible that the results for $s$-wave (nodeless) and $d$-wave (nodal) superconductors  are applicable for other 2d states of matter with, respectively, a fully gapped or nodal quasiparticle spectrum. 

\section{Strategy}
The starting assumption is that the ground-state in the 2d limit  is a superconductor.  It is well known that such a system has a finite temperature power-law phase which is well described in terms of a non-linear sigma model describing the phase modes.  By analogy, the same behavior must occur on any large cylinder at $T=0$ with ${L_{\perp}}$ playing the same role in the quantum physics as $\beta=1/T$ does in the thermal \footnote{The finite temperature problem in 2d  can be formulated in terms of an imaginary-time path integral on a Euclidean space with an emergent rotational (Lorentz) invariance and periodic boundary conditions in the time-direction.  The $T=0$ problem on a cylinder with periodic boundary conditions in the $y$ direction thus is formally similar if we identify $y$ with time.}.
Thus, in Sec. \ref{sec:collective}, we use a $\mathrm{U}(1)$ non-linear sigma model to derive expressions for the long-distance behavior of the superconducting and CDW correlations in terms of the superfluid stiffness and phase mode velocity of the 2d system.  At a more microscopic level,
so long as ${L_{\perp}}$ is sufficiently large, the ground state can be well described 
from a starting point of BCS theory. 
In Secs. \ref{sec:BCS_general}, \ref{sec:BCS_s-wave}, and \ref{sec:BCS_d-wave},
we have explicitly worked out finite-${L_{\perp}}$ BCS theory for a class of lattice models.  The finite-size scaling properties of various correlations that depend on the quasi-particle spectrum are determined.

\section{Collective correlations}
\label{sec:collective}

To address the physics of the $\text{U}(1)$ Goldstone mode, i.e. the phase of the superconducting order parameter, we take as our starting point  the effective action
\begin{equation}
    S = 
    \int \dd\tau \int \dd x \int_0^{L_{\perp}} \dd y 
    \
    \frac{\rho_s}{2}  \bqty{ (\nabla \theta)^2
    + \frac{1}{v^2} (\partial_{\tau} \theta)^2
    }.
    \label{eq:sc_action}
\end{equation}
Here, the field $\theta$ is the phase-angle of the order parameter, the parameters $\rho_s$ and $v$ are the superfluid stiffness and mode velocity, respectively, and periodic boundary conditions are imposed in $y$. For simplicity we assume an isotropic superfluid stiffness. (The anisotropic case reduces to the isotropic one by re-scaling coordinates.)
It is implicit that there is an ultra-violet cuttoff provided by the BCS coherence length.

Note that all transverse phase fluctuations are gapped. They can be integrated out, leaving behind the action for a standard 1d superfluid, i.e. 
a Luther-Emery liquid.

\subsection{Superconducting correlations}

The superconducting order parameter is $\Phi(x,y) =  \Phi_0 e^{i\theta(x,y)}$. As shown in the Appendix~\ref{app:sc_corr}, the superconducting correlation function for $\abs{x} \gg {L_{\perp}} \gg 1$ is 
\begin{equation}
   \ev{\Phi^*(x,y) \Phi(0)} 
   \sim \abs{\ev{\Phi}_{2d}}^2
     \pqty{\dfrac{A {L_{\perp}}}{\abs{x}}}^{\frac{v}{2\pi \rho_s {L_{\perp}}}}. 
    \label{eq:sc_corr}
\end{equation}
(At shorter distances the correlation function behaves as in the 2d limit.)
Here $ \ev{\Phi}_{2d}$ is the expectation value of the order parameter evaluated in the 2d limit and $A$ is a constant.  
The explicit calculation in Appendix~\ref{app:sc_corr} gives the universal value $A \approx 1.12$. 
The important point is that the power-law exponent vanishes as ${L_{\perp}} \to \infty$. 
To illustrate the significance of this, note that the distance, $\ell_x$, over which the correlation function decays to $1/e$ of its 2d value is exponentially long;  $\ell_x = A{L_{\perp}} e^{2\pi \rho_s {L_{\perp}}/v}$. (Note that $\ell_x$ is not a correlation length.)

We note that in the DMRG study \cite{Chatterjee_2020}, power-law superconducting correlations with an exponent clearly scaling as $1/L_{\perp}$ were observed in a certain model, and used to identify a superconducting phase in the 2d limit.

\subsection{CDW correlations}
It is well known that a one-dimensional superfluid exhibits not only quasi-long range superconducting order but also quasi-long range CDW order: The density-density correlation function at long distances is the sum of two pieces -- a non-oscillating, hydrodynamic piece that decays as $1/x^2$, and a CDW-like piece with asymptotic form
\begin{equation}
    (\ev{n(x)n(0)})_{cdw} \sim 
    \cos(Q x + \alpha) 
    \ \abs{x}^{-K}
    \label{eq:asymptotic_cdw}
\end{equation}
where the exponent $K$ is the inverse of the exponent characterizing the power law for superconducting correlations, $Q=\pi\nu$ with $\nu$ the total number of electrons per unit cell, $\alpha$ is some phase, and the amplitude depends upon microscopic details \cite{Wen_2010}.
For the cylinder problem at hand, $\nu = {L_{\perp}} \ev{n}$ where $n$ is the number of electrons per site, and $K = 2\pi \rho_s {L_{\perp}}/v$.

CDW correlations with this asymptotic form have been found in numerical and analytical studies on small-${L_{\perp}}$ Luther-Emery liquids \cite{Balents_Fisher_1996,Gannot_Jiang_Kivelson_2020,Noack1994,Noack_White_Scalapino_1996}. However, it is easy to see that with increasing ${L_{\perp}}$, this asymptotic form rapidly becomes nearly impossible to observe. Out to distances $\sim {L_{\perp}}$, all charge density correlations behave as they would in the two-dimensional limit: any oscillating piece (e.g. Friedel oscillations) exhibits wave-vectors that  reflect the structure of the underlying  Fermi surface, and possibly details of the gap structure. It is only at longer distances that power-law correlations with wave-vector $Q=\pi {L_{\perp}} \langle n\rangle $ appear, but they fall with a large exponent,  $K = 2\pi \rho_s {L_{\perp}}$, and typically have a small amplitude. 
(This may explain the failure to detect the expected CDW correlations in certain DMRG studies of relatively wide ladders \cite{Jiang_Kivelson_2022}.)

More generally, any cylinder system such that $\nu = L_{\perp} \ev{n}$ is not an even integer possesses either true or quasi long-range CDW order at wavevector $Q = \pi \nu$. This follows from the generalized Lieb-Schultz-Mattis theorem \cite{Yamanaka_Oshikawa_Affleck_1997}. As in the superconducting case described above, if the 2d limit has only short-range CDW correlations then the order at wavevector $Q$ typically disappears quite rapidly as a function of increasing $L_{\perp}$. For instance, fractional quantum hall states and $\mathbb{Z}_2$ spin liquids on cylinders exhibit density oscillations at wavevector $Q$ whose amplitude decays exponentially with $L_{\perp}$ \cite{Yao_Kivelson_2012,Jiang_Yao_Balents_2012,Seidel_et_al_2005,Strinati_et_al_2019}.

\section{BCS analysis of lattice models}
\label{sec:BCS_general}
We next turn to a finite-${L_{\perp}}$ BCS analysis of lattice models with attractive microscopic interactions. They are valid models in their own right, but can also be considered as effective models for systems where the pairing interaction arises from more complicated physics. In this section we will present the relevant models and the general structure of the theory. Specific applications to $s$-wave and $d$-wave superconductors -- which have qualitatively different finite-size scaling due to different quasiparticle spectra -- are given in Sections \ref{sec:BCS_s-wave} and \ref{sec:BCS_d-wave}.

\subsection{Models}
 The Hamiltonian is 
\begin{multline}
    H = \sum_{\substack{ \vb{r} \in \mathbb{Z}^2
    \\
    \vb{r}' \in S({L_{\perp}})}} \bigg[\sum_{\sigma}t(\vb{r}-\vb{r}') c^{\dg}_{\vb{r} \sigma}c_{\vb{r}'\sigma}  
     \\ + \frac{1}{2}\sum_{\sigma \sigma'}  V_{\sigma
     _1\sigma_2;\sigma_3,\sigma_4}(\vb{r}- \vb{r}') c^{\dg}_{\vb{r}\sigma_1}c^{\dg}_{\vb{r}'\sigma_2}c_{\vb{r}'\sigma_4}c_{\vb{r}\sigma_3} \bigg],
     \label{eq:hamiltonian}
\end{multline}
where $c^{\dg}_{\vb{r}\sigma}$ creates an electron at site $\vb{r} = (x,y)$ with spin polarization $\sigma$ and $S({L_{\perp}}) = \mathbb{Z} \times \{0,1,\ldots, {L_{\perp}}-1\}$ is a strip of width ${L_{\perp}}$. The functions $t$ and $V$ are real, finite ranged, and even in $\vb{r}$, and the spin structure on $V$ is rotationally invariant. Finally, since one of the most useful tools in studying the approach to the thermodynamic limit is to explore the sensitivity to boundary conditions, we make the identification
\begin{equation}
    c_{(x,y+{L_{\perp}})\sigma} = e^{i\sigma \phi}c_{(x,y)\sigma}
    \label{eq:identification}
\end{equation}
on the fields, implementing spin-twisted periodic boundary conditions in $y$. [We let $\sigma = 1$ ($-1)$ for spin up (down).] This corresponds to the presence of an Aharanov-Bohm flux $\phi$ through the interior of the cylinder, with opposite charge for opposite spins. Note that 
these boundary conditions preserve time reversal symmetry for arbitrary $\phi$,
and that 
any $\phi$-dependence must vanish as ${L_{\perp}} \to \infty$.

\subsection{BCS theory on a cylinder}
If we assume that the 2d limit has a spin singlet gap function, then at sufficiently large ${L_{\perp}}$ we can consider a BCS trial Hamiltonian of the form
\begin{multline}
    H_{\text{trial}} = \sum_{\substack{ \vb{r} \in \mathbb{Z}^2
    \\
    \vb{r}' \in S({L_{\perp}})}} \bigg[ \tau(\vb{r} - \vb{r}') c^{\dg}_{\vb{r} \up}c_{\vb{r}'\up}  +\tau^*(\vb{r} - \vb{r}') c^{\dg}_{\vb{r} \down}c_{\vb{r}'\down} \\
    + \pqty{ \tilde\Delta(\vb{r}-\vb{r}')c^{\dg}_{\vb{r} \up}c^{\dg}_{\vb{r}' \down} + \hc} \bigg].
    \label{eq:trial_ham}
\end{multline} 
where the fields satisfy the 
boundary conditions  Eq.~(\ref{eq:identification}). 

Since the flux preserves
spin rotation about $z$, we can continue to assume zero-momentum, $S_z = 0$ pairing. However, since  
 non-zero twist angle, $\phi\neq 0$,  breaks symmetry under inversion and spin rotation about any axis besides $z$, the gap does not have definite parity or $S^2$: mixed in with the dominant even-parity singlet component is an odd-parity triplet component, whose amplitude vanishes as ${L_{\perp}} \to \infty$; likewise $\tau$, in principle, contains a small odd-parity piece.

Before discussing the self-consistency equations imposed on $\tilde{\Delta}$ and $\tau$, we briefly review the properties of the trial Hamiltonian at arbitrary $\tilde{\Delta}$ and $\tau$. These are worked out in Appendix~\ref{app:trial_ham}. The energy of a quasiparticle with spin polarization $\sigma$ and momentum $\vb{k}$ is $E(\sigma \vb{k})$, where
\begin{equation}
    E(\vb{k}) = \sqrt{\varepsilon^2(\vb{k}) + \Delta(\vb{k})^2}.
\end{equation}
Here $\varepsilon(\vb{k}) = \sum_{\vb{r}} e^{-i \vb{k} \vdot \vb{r}} \tau(\vb{r}) - \mu$ is the band energy minus chemical potential and $\Delta(\vb{k}) = -\sum_{\vb{r}}e^{-i \vb{k} \vdot \vb{r}} \tilde \Delta(\vb{r})$, which we choose without loss of generality to be real. Owing to the cylinder geometry and the flux, $\vb{k}$ is restricted to lie along momentum space slices satisfying 
\begin{equation}
    k_y = (2\pi n + \sigma \phi)/{L_{\perp}}, \quad n \in \mathbb{Z}.
    \label{eq:allowed_momenta}
\end{equation}
Finally, the expectation value of any operator can be computed, via Wick's theorem, in terms of the basic correlation functions
\begin{align}
    &F(\vb{r} -\vb{r}') =  \langle c_{\vb{r}'\down}c_{\vb{r}\up}\rangle_{\text{trial}}, \label{eq:F_def} \\
    &G(\vb{r} - \vb{r}') = \langle  c^{\dg}_{\vb{r}' \up}c_{\vb{r}\up} \rangle_{\text{trial}} \label{eq:G_def}.
\end{align}
These depend parametrically upon $\Delta$, ${L_{\perp}}$, and $\phi$, but when there is no potential for confusion we will leave the dependence implicit.

Now we turn to the self-consistency equations, derived in Appendix~\ref{app:self_consistency}. In principle, both the gap $\tilde{\Delta}$ and the hopping $\tau$ must be
determined self-consistently at each ${L_{\perp}}$ and $\phi$. Here, we present a simplified discussion in which we fix $\tau$ to its value in the 2d limit, which gives qualitatively the same result as other schemes \footnote{To allow the fixed $\tau$ to be real at arbitrary $\phi$, we henceforth generalize Eq.~(\ref{eq:hamiltonian}) to allow for non-real $t$, analogous to the hopping in Eq.~(\ref{eq:trial_ham})}. The self-consistency equation for $\tilde{\Delta}$ is
\begin{align}
    \tilde{\Delta}(\vb{r}) = &\frac{V_s(\vb{r})}{2}\bqty{F(\vb{r}) 
    + F(-\vb{r})}
    + \frac{V_t(\vb{r})}{2} \bqty{F(\vb{r}) - F(-\vb{r})},
    \label{eq:gap_eqn_main}
\end{align}
where $V_s(\vb{r})$ and $V_t(\vb{r})$ are the singlet and triplet eigenvalues of the interaction at separation $\vb{r}$. 

Within the BCS theory presented here, we have solved analytically for the leading order finite-size corrections to various quantities. The systematic framework for carrying out these calculations is described in Appendix~\ref{app:finite_circumference_perturbation}. We now proceed to present the results, along with corroborating numerical examples. In what follows, a subscript
$2d$ on any quantity denotes evaluation in the 2d limit, ${L_{\perp}} \to \infty$.

\section{BCS analysis of a fully gapped superconductor}
\label{sec:BCS_s-wave}

As the first application of the BCS formalism, we consider a system which we stipulate has a fully gapped (s-wave) superconducting ground-state in the 2d limit, i.e. such that $E_{2d}(\vb{k}) > 0$ for all $\vb{k}$. Such a state has exponentially falling single-particle and spin-spin correlation functions. The calculational details for this section may be found in Appendix~\ref{app:fully_gapped}.

\subsection{Corrections to expectation values}

We start with the finite size scaling for ground state expectation values of local operators,
as well as of the gap function which inherits ${L_{\perp}}$ dependence from the self consistency equation. It is easy to see that finite size corrections to these quantities are related to correlation functions in the 2d limit, evaluated at separation ${L_{\perp}}\vu{y}$. For any fully gapped superconductor, the large-${L_{\perp}}$ form of the latter is
\begin{equation}
    F_{\text{2d}}({L_{\perp}}\vu{y}) \sim G_{2d}({L_{\perp}}\vu{y})
    \sim
    \cos({Q_{\perp}} {L_{\perp}} + \alpha)e^{-{L_{\perp}}/{\xi_{\perp}}}
\end{equation}
for some correlation length ${\xi_{\perp}}$, wavevector ${Q_{\perp}}$, and phase shift $\alpha$. 
(For simplicity, power law prefactors 
will be omitted here and 
henceforth wherever we find an exponential decay.  It is also typically the case that $\alpha$ takes distinct values for $F$ and $G$, and indeed takes on unique values in each asymptotic relation below.) 
By symmetry, the expectation of any odd parity operator vanishes in the 2d limit (${L_{\perp}}\to\infty$), and still does so for finite ${L_{\perp}}$ as well when $\phi=0$ or $\pi$.  However,  the expectation value of any even parity operator, $A({L_{\perp}},\phi)$, approaches its value in the 2d limit as
\begin{equation}
    \left[A({L_{\perp}},\phi) - A_{2d} \right]\sim
    e^{-{L_{\perp}}/{\xi_{\perp}}}
    \cos(\phi)\cos({Q_{\perp}} {L_{\perp}} + \alpha) 
   \label{eq:fully_gapped_correction}
\end{equation}
with the same ${\xi_{\perp}}$ and ${Q_{\perp}}$
that characterize the 2d limit correlation functions.
This means that for the special values of the flux,
$\phi = \pm \pi/2$, the leading order corrections to even quantities vanish, and 
\begin{equation}
   \left[ A({L_{\perp}},
    \pm \pi/2) - A_{2d}\right] \sim e^{-2{L_{\perp}}/{\xi_{\perp}}}
   \cos(2Q_{\perp} L_{\perp} + \alpha)
    \label{eq:fully_gapped_correction_special_flux}
\end{equation}

It is important to note that not every observable is required to converge exponentially fast, because some observables are not expectation values of local operators. For example, consider the excitation energy $\mathcal{E}$ to the lowest energy quasiparticle at fixed flux, 
equal to
\begin{equation}
    \mathcal{E}({L_{\perp}},\phi) = \min_{\text{allowed $\vb{k}$}}E(\vb{k}),
\end{equation}
where ``allowed $\vb{k}$'' refers to 
values consistent with the relevant boundary conditions, Eq.~(\ref{eq:allowed_momenta}). 
For given $\tilde{\Delta}$,
$E(\vb{k})$ is well-defined at all momenta,
so on the basis of the the preceding discussion,
its minimum value over all $\vb{k}$ would exhibit only exponentially small finite size variations, $\mathcal{O}(e^{-{L_{\perp}}/{\xi_{\perp}}})$. 
However, the 2d momentum at which $E(\vb{k})$ is minimal
typically
lies a distance $\mathcal{O}(1/{L_{\perp}})$ from the nearest allowed momentum, leading to
\begin{equation}
    \mathcal{E}({L_{\perp}},\phi) - \mathcal{E}_{2d} = \mathcal{O}(1/{L_{\perp}^2}).
\end{equation}

Faster convergence to the 2d limit 
can be achieved by continuously cycling $\phi$ from $0$ to $2\pi$ at fixed ${L_{\perp}}$. The allowed momentum slices then shift upward
in lock-step by an amount $2\pi/{L_{\perp}}$ (the spacing between slices). At some point in the process, the approximate minimum hits one of the slices; thus
\begin{equation}
    \min_{\phi} \left[\mathcal{E}({L_{\perp}},\phi)\right]  - \mathcal{E}_{2d} = \mathcal{O}(
    e^{-{L_{\perp}}/{\xi_{\perp}}}).
    \label{eq:fully_gapped_minimum}
\end{equation}

\subsection{
Anisotropy and the approach to the 2d limit}

Although introducing lattice anisotropy changes the point-group symmetry, a fully gapped phase must evolve smoothly.  Specifically, 
if a 2d system with a lattice $C_4$ symmetry has an $s$-wave superconducting ground state,
the same will be true if the hopping matrix elements in the $x$ and $y$ directions are made somewhat unequal. Moreover, it is clear that if the hopping in the $y$ direction is reduced, this should reduce the correlation length ${\xi_{\perp}}$ in that direction and speed up the approach to the thermodynamic limit as a function of increasing ${L_{\perp}}$.

\begin{figure}[h]
\centering
\includegraphics[width=0.9\linewidth]{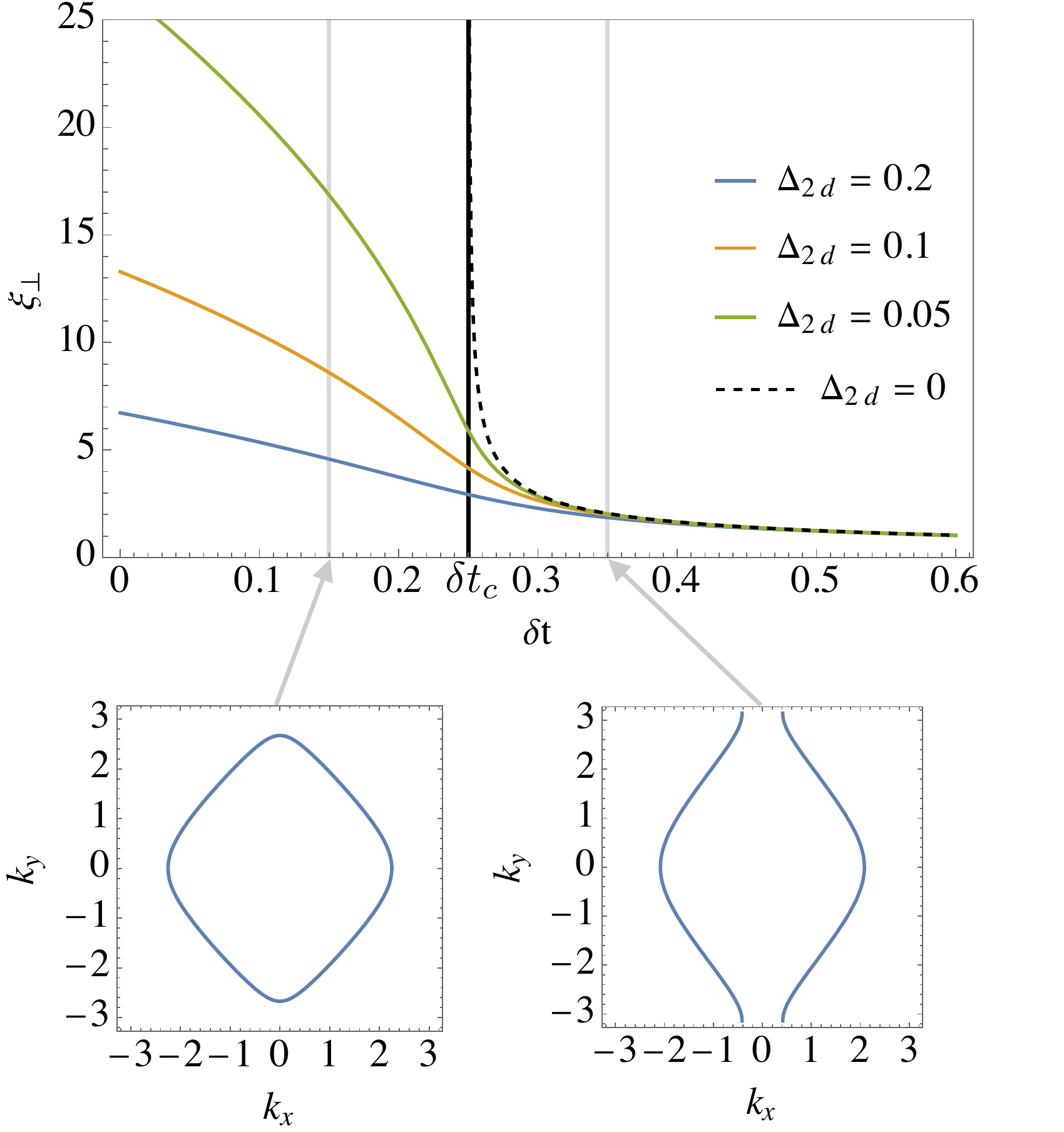}
\caption{Top: Plot of ${\xi_{\perp}}$  as a function of $\delta t = t_x - t_y$, for several fixed values of $\Delta_{2d}$, for the model defined by Eq.~(\ref{eq:simple_model}). The remaining parameters are fixed as $(t_x + t_y)/2 = 1$, $\mu = -0.5$. The dark vertical line at $\delta t_c = 0.25$ separates closed and open Fermi surfaces. For the Fermi liquid case $\Delta_{2d} = 0$ (dashed line), ${\xi_{\perp}}$ is finite only for $\delta t > \delta t_c$.
Bottom: representative closed and open Fermi surfaces for $\delta t = 0.15$ and $\delta t = 0.35$, respectively.
\label{fig:xi_y}}
\end{figure}

However, the degree to which $\xi_{\perp}$ is reduced by anisotropy can be surprisingly large. As a warm-up to the superconductor problem, let us consider a 2d Fermi liquid. In the isotropic case, correlations fall off algebraically in all directions. So long as the  Fermi surface remains closed, the same is true upon introducing anisotropy. However, if anisotropy is sufficiently strong that the Fermi surface is open, there can be (and typically will be) directions along which correlations decay exponentially, even at $T= 0$. Specifically, correlations decay exponentially in any direction $\vu{e}$ along which electrons on the Fermi surface are kinematically forbidden from propagating, i.e. any $\vu{e}$ which is not is parallel to the Fermi velocity $\vb{v}_F$ at any point on the Fermi surface. Such directions form a fan about $\pm \vu{y}$ -- see for example the open Fermi surface shown in the lower right panel of Fig.~\ref{fig:xi_y}. The upshot is that for a Fermi liquid, introducing sufficient hopping anisotropy to make the Fermi surface open reduces $\xi_{\perp}$ from $\infty$ to something on the order of a lattice spacing.

The effect is similarly dramatic for
a superconductor with a gap function that is small as compared to the bandwidth.
To illustrate this point, we have treated 
the 2d state described by a trial Hamiltonian with hopping and gap functions corresponding to
\begin{subequations}
\label{eq:simple_model}
\begin{align}
    & \varepsilon(\vb{k}) = -2 t_x \cos(k_x)- 2 t_y \cos(k_y) - \mu \label{eq:nn_dispersion} \\
    & \Delta_{2d}(\vb{k}) = \Delta_{2d} \label{eq:uniform gap}
\end{align}
\end{subequations}
for constant $\Delta_{2d}$. It is straightforward to find ${\xi_{\perp}}$ in this model -- see Appendix \ref{app:xi_perp}. The result is depicted in Fig.~\ref{fig:xi_y}. The top panel illustrates how ${\xi_{\perp}}$  evolves as a function of the anisotropy parameter $\delta t = t_x - t_y$, and the bottom panel shows representative Fermi surfaces. There is a critical $ \delta t_c$ with the following property:
\begin{itemize}
    \item For $\delta t < \delta t_c$: The underlying Fermi surface is closed in $k_y$, $\vb{v}_F$ points along $\vu{y}$ at the top of the Fermi surface, and ${\xi_{\perp}} \propto 1/\Delta_{2d}$.
    \item For $\delta t > \delta t_c$: The underlying Fermi surface is open in $k_y$, $\vb{v}_F$ never points along $\vu{y}$, and ${\xi_{\perp}}$ is on the order of a lattice spacing.
\end{itemize}
(To be explicit, $\delta t_c$ is the critical value of $\delta t$ separating closed and open Fermi surfaces in the $\Delta_{2d} = 0$ system.)

By Eq.~(\ref{eq:fully_gapped_correction}), it is ${\xi_{\perp}}$ which controls the approach to the 2d limit of various properties of the system. Therefore, for $s$-wave superconductors characterized by an emergent correlation length $\xi_{2d} \equiv v_F/\Delta_{2d} \gg 1$ (where $v_F$ and $\Delta_{2d}$ are typical values of the Fermi velocity and gap, respectively) we arrive at the following conclusion: While in the isotropic case the 2d limit is approached only when $L_{\perp}  \gg \xi_{2d}$, if anisotropy is sufficiently strong that the Fermi surface is open, 2d physics is apparent when ${L_{\perp} \gg 1}$, even if $\xi_{2d} \gg L_{\perp}$.

\subsection{Numerical examples}

Here, 
we numerically study the 
the crossover from small to large ${L_{\perp}}$. We consider a model with an onsite attraction, $V_s(\vb{r}) = V\delta_{\vb{r},\vb{0}}$ for some $V < 0$  and $V_{t}(\vb{r}) = 0$. As we discussed above, rather than fixing $t$ we fix $\tau$, which we now specify corresponds to the nearest neighbor dispersion $\varepsilon(\vb{k}) = -2 t_x \cos(k_x)- 2 t_y \cos(k_y) - \mu$. 
(Note that in any case, $\tau$ and $t$ for an onsite interaction would differ only by a trivial onsite energy.) 
At all ${L_{\perp}}$, the gap is purely onsite: $\tilde{\Delta}(\vb{r}) = \Delta \delta_{\vb{r},0}$ where $\Delta = \Delta({L_{\perp}}, \phi)$.

 We present data for two observables: the gap, and the superfluid stiffness down the length of the cylinder,
\begin{equation}
\rho_{s}^{x} \equiv \left. \frac{1}{\Omega}\pdv[2]{E}{q_x}  \right|_{q_x = 0},
\end{equation}
where $E/\Omega$ is the ground state energy per unit area in the presence of a phase-twisted gap, i.e. under the replacement $\Delta c^{\dg}_{\vb{r} \up}c^{\dg}_{\vb{r} \down} \to e^{i q_x x} \Delta c^{\dg}_{\vb{r} \up}c^{\dg}_{\vb{r} \down}$ in the trial Hamiltonian \cite{Fisher_Barber_David_1973,Scalapino_White_Zhang_1993}. $\rho_{s}^{x}$ in this model reduces to the ground state expectation value of the negative of the kinetic energy density associated with $x$-hopping.

\begin{figure*}
\centering
\includegraphics[width=\linewidth]{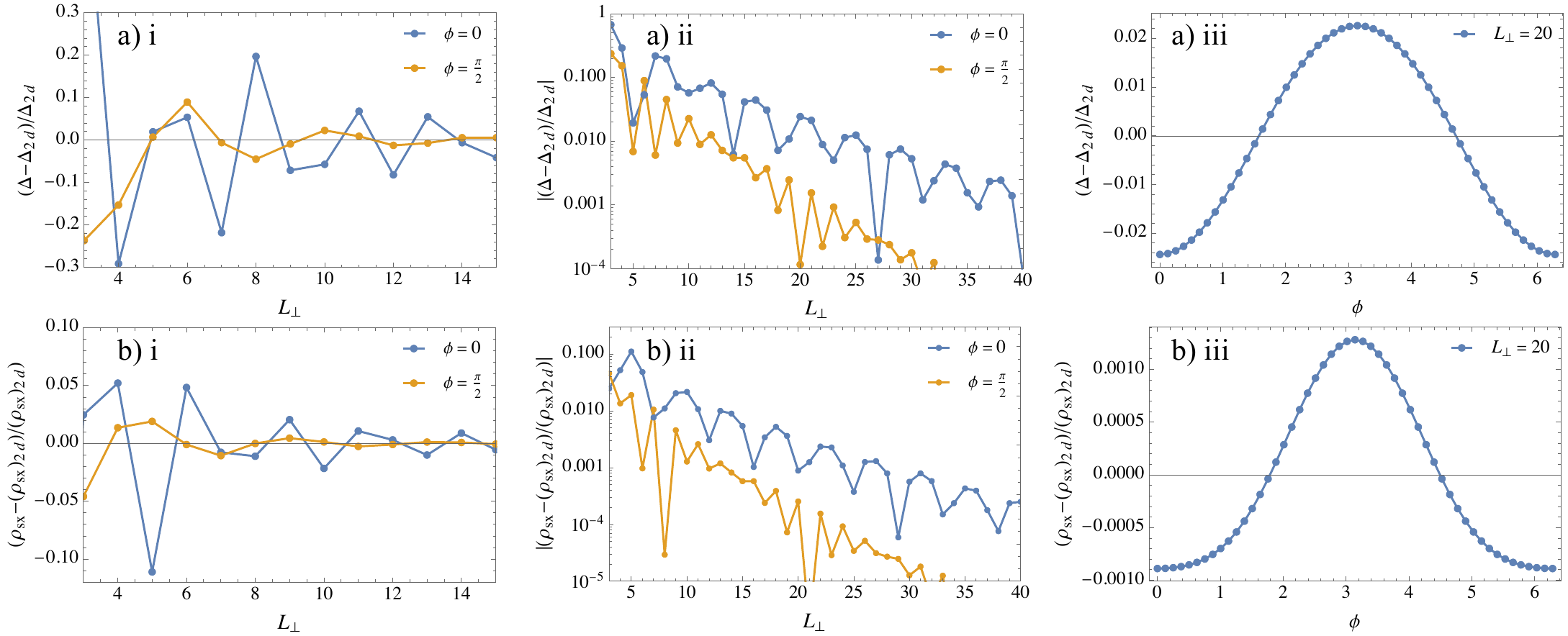}
\caption{Results of BCS calculations for $t_x = t_y = 1$, $\mu = -0.5$, and $V = -1.5$. The two-dimensional gap for these parameters is $\Delta_{2d} \approx 0.126$. Panels a)i-iii in the top row depict the relative finite-${L_{\perp}}$ error in the gap, and panels b)i-iii in the bottom row are the same but for the superfluid stiffness. a)i and b)i show the ${L_{\perp}}$-dependence observables at $\phi = 0$ and the special flux value $\phi = \pi/2$; a)ii and b)ii show the same thing but on a log scale and up to larger ${L_{\perp}}$; a)iii) and b)iii show the $\phi$-dependence at ${L_{\perp}} = 20$.
\label{fig:exp_decay_flux}}
\end{figure*}

First, we illustrate the dependence of observables upon ${L_{\perp}}$ and $\phi$.
Fig.~\ref{fig:exp_decay_flux} shows representative results for an isotropic model. At fixed $\phi$, we
see exponentially decaying oscillations for the finite-${L_{\perp}}$ corrections to both $\Delta$ and $\rho_{s}^{x}$. We also see that these corrections decay at an asymptotically faster rate for $\phi = \pi/2$ than $\phi = 0$, and that even for moderate ${L_{\perp}}$ setting $\phi = \pi/2$ substantially mitigates finite size errors. 
We also see the 
$\cos(\phi)$-dependence to corrections predicted for large fixed ${L_{\perp}}$.

Next, we illustrate how a small amount of hopping anisotropy favoring $x$-hopping over $y$-hopping can substantially speed up convergence to the two-dimensional limit. To isolate the interplay between cylinder geometry and hopping anisotropy from changes to the 2d state being approached, we consider two models related by swapping $t_x \leftrightarrow t_y$, i.e. by a $90^{\circ}$ rotation of the hopping parameters. We chose the hoppings so that the first model has a Fermi surface open along $k_x$ and closed along $k_y$, and the second has a Fermi surface closed along $k_x$ and open along $k_y$.

\begin{figure}
\centering
\includegraphics[width=0.95\linewidth]{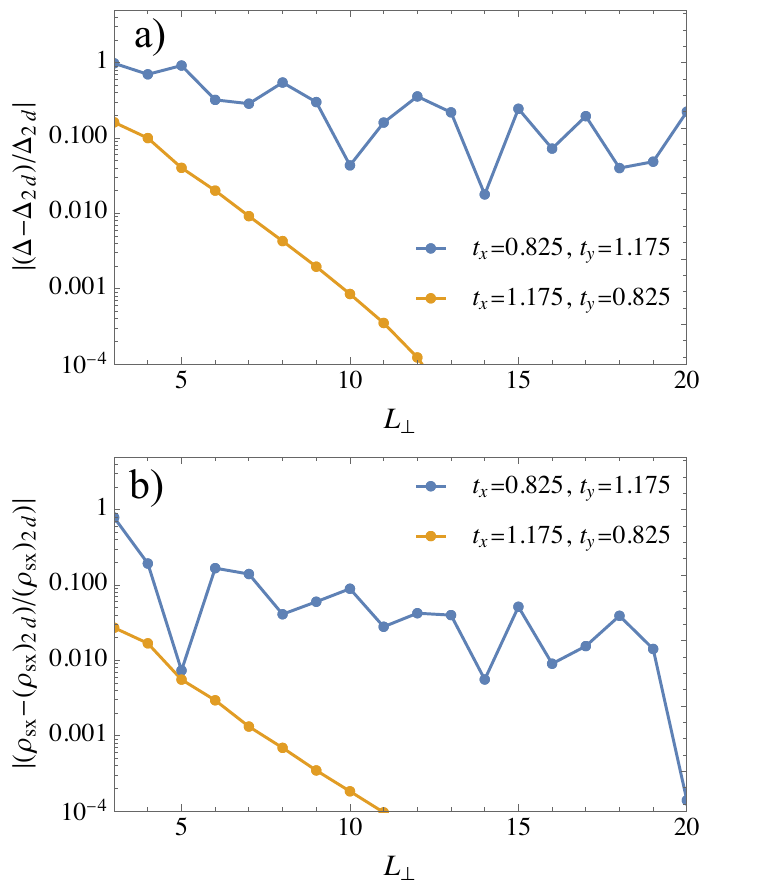}
\caption{
Results of BCS calculations for two models. In the first, which corresponds to the blue plots, $t_x = 0.825, t_y = 1.125$ which gives a Fermi surface closed in $k_y$ and open in $k_x$. In the second, which corresponds to the orange plots, $t_x = 1.125, t_y = 0.825$ which gives a Fermi surface open in $k_y$ and closed in $k_x$. In both cases we take $\mu = -0.5$ and $V = -1.25$, and we fix $\phi = 0$ on the cylinders. The two models have the same 2d limit gap $\Delta_{2d} \approx 0.055$. Panel a) shows the finite-${L_{\perp}}$ correction to the gap, and panel b) shows the same for the superfluid stiffness.
\label{fig:anisotropy_influence}}
\end{figure}

These two models have the same gap in the 2d limit. However, when $\Delta_{2d}$ is small, the gap approaches the 2d limit very differently in the two cases. Fig.~\ref{fig:anisotropy_influence}a) shows results for a case where $\Delta_{2d} \approx 0.055$. At moderate ${L_{\perp}}$, $\Delta$ in the model with a Fermi surface closed in $k_y$ suffers large relative errors, whereas $\Delta$ in the model with a Fermi surface open in $k_y$ is fully converged. We also present analogous results for $\rho_{s}^x$ in Fig.~\ref{fig:anisotropy_influence}b), though it should be noted in this case that $\rho_{s}^x$ approaches a different 2d limit in the two models.

\section{BCS analysis of a nodal superconductor}
\label{sec:BCS_d-wave}

Next, we study the case 
in which the cylinder in question approaches a nodal superconductor in the 2d limit. Such a state is characterized by a nodal quasiparticle spectrum, i.e. points $\vb{Q}_{2d}$ on the Fermi surface at which the gap function changes sign. In the vicinity of such points the quasiparticle spectrum takes a gapless Dirac form.

For the sake of definiteness, we assume that in the 2d limit, there exist exactly four nodal momenta for a given spin polarization: the pair $\pm \vb{Q} = \pm (Q_{\parallel,2d}, Q_{\perp,2d})$, the members of which are related by spatial inversion, and the pair $\pm (-Q_{\parallel,2d}, Q_{\perp,2d})$ related to the first pair by an assumed reflection symmetry in $x$.

The calculational details for this section may be found in Appendix \ref{app:nodal}.

\subsection{Corrections to expectation values}
Similar to the case of a fully gapped superconductor, finite size corrections to expectation values in a nodal superconductor are related to correlation functions in the 2d limit at separation ${L_{\perp}} \vu{y}$. The latter take the large-${L_{\perp}}$ form
\begin{equation}
    F_{2d}({L_{\perp}} \vu{y}) \sim G_{2d}({L_{\perp}} \vu{y}) \sim \sin({L_{\perp}} Q_{\perp,2d})/{L_{\perp}^2}
\end{equation}
However, since the 2d correlations are long-ranged in the present case, their precise relation to the finite size corrections is somewhat complicated. We find that expectation values $A$ of even-parity operators satisfy
\begin{multline}
     A({L_{\perp}}, \phi) - A_{2d} \\ \sim \frac{1}{{L_{\perp}^2}} \bqty{\text{Cl}_2\pqty{{L_{\perp}} Q_{\perp,2d} - \phi} + \text{Cl}_2\pqty{{L_{\perp}} Q_{\perp,2d} + \phi} } \\
     \label{eq:nodal_expect_correction}
\end{multline}
where
\begin{equation}
    \text{Cl}_2(\theta) \equiv \sum_{m=1}^{\infty} \sin(m \theta)/m^2 \label{eq:clausen_def}
\end{equation}
is conventionally known as a Clausen function of order 2 and is odd, $2\pi$-periodic, and has a logarithmically diverging derivative at $\theta = 2\pi n$ for $n \in \mathbb{Z}$. Note that in Eq.~(\ref{eq:nodal_expect_correction}), the arguments to the Clausen functions are ${L_{\perp}}$ times the separation along the $y$-axis between a given 2d nodal point and any allowed momentum slice, modulo $2\pi$. 

In contrast with the fully gapped case,  no single twist angle can eliminate the leading finite size correction. However, $\text{Cl}_2$ has zero average value (being odd and periodic) so averaging uniformly over twist angles does eliminate the leading correction:
\begin{equation}
    \ev{A({L_{\perp}}, \phi)}_{\phi} - A_{2d} = \order{1/L_{\perp}^3},
    \label{eq:nodal_phi_average}
\end{equation}
where $\ev{\hspace{3 pt}}_{\phi}$ denotes a uniform average over $\phi$.

\subsection{The quasiparticle gap}
The quasiparticle energy near a node is proportional to the distance from the node. Generically, each node lies an $\order{1/{L_{\perp}}}$ distance away from the nearest allowed momentum slice. Thus 
\begin{equation}
    \mathcal{E}({L_{\perp}},\phi) = \order{1/{L_{\perp}}}.
\end{equation}
By varying the twist angle $\phi$, it is possible to tune the distance between the allowed momenta slices and the nodes. The quasiparticle gap thus vanishes substantially faster with ${L_{\perp}}$ when it is first minimized over $\phi$, an observation that was 
used in Refs.~\cite{He_et_al_2017, Hu_et_al_2019} in the spin liquid context. We find two possibilities:
\begin{enumerate}
    \item $\min_{\phi} \mathcal{E}({L_{\perp}},\phi) = 0$
    \item $\min_{\phi} \mathcal{E}({L_{\perp}},\phi) = \order{e^{-{L_{\perp}}/\ell_n}}$ for some $\ell_n > 0$
\end{enumerate}
In the first case, momentum slices collide with the nodes upon continuously increasing the twist angle.
In the second case, the collision is avoided -- to remain in the lowest energy state, the node jumps across any 
momentum slice that would intersect the node. The exact criterion distinguishing the two cases depends upon microscopic details and is provided in Appendix~\ref{app:qp_gap}. However, we note here that one situation in which the second case occurs is if the interaction is fully attractive, in the sense that $V_s(\vb{r}), V_t(\vb{r}) \le 0$ for all $\vb{r}$.

\subsection{Numerical examples}
\label{sec:nodal_numerical}
We now numerically study 
the crossover from small to large ${L_{\perp}}$. We work with the $C_4$ symmetric nearest neighbor dispersion  $\varepsilon(\vb{k}) = -2 \cos(k_x) - 2\cos(k_y) - \mu$ and the nearest neighbor interaction 
\begin{equation}
    V_{s/t}(\vb{r}) = V_{s/t} \sum_{\vu{e} \in \{\pm \vu{x}, \pm \vu{y} \}} \delta_{\vb{r},\vu{e}}.
\end{equation}
In the 2d limit, the 
solution to the self-consistency equation has the nearest neighbor $d$-wave gap function 
\begin{equation}
    \Delta_{2d}(\vb{k}) = \Delta_{2d} (\cos(k_x) - \cos(k_y)),
    \label{eq:2d_dwave}
\end{equation}
and corresponding nodes in the spectrum at $\pm(Q,Q)$, $\pm(-Q,Q)$. 
in the appropriate range of parameters.

At finite ${L_{\perp}}$, the gap function generically contains nearest neighbor $s$ and $p$-wave components, taking the form
\begin{multline}
    \Delta(\vb{k}) = \Delta^d (\cos(k_x) -\cos(k_y))  \\ + \Delta^s (\cos(k_x) +  \cos(k_y)) + \Delta^p \sqrt{2} \sin(k_y).
\end{multline}
The fact that this model at arbitrary ${L_{\perp}}$ and $\phi$ preserves mirror symmetry in $x$ means 
that there is no  
gap component proportional to $\sin(k_x)$. 

\begin{figure*}
\centering
\includegraphics[width=0.75\linewidth]{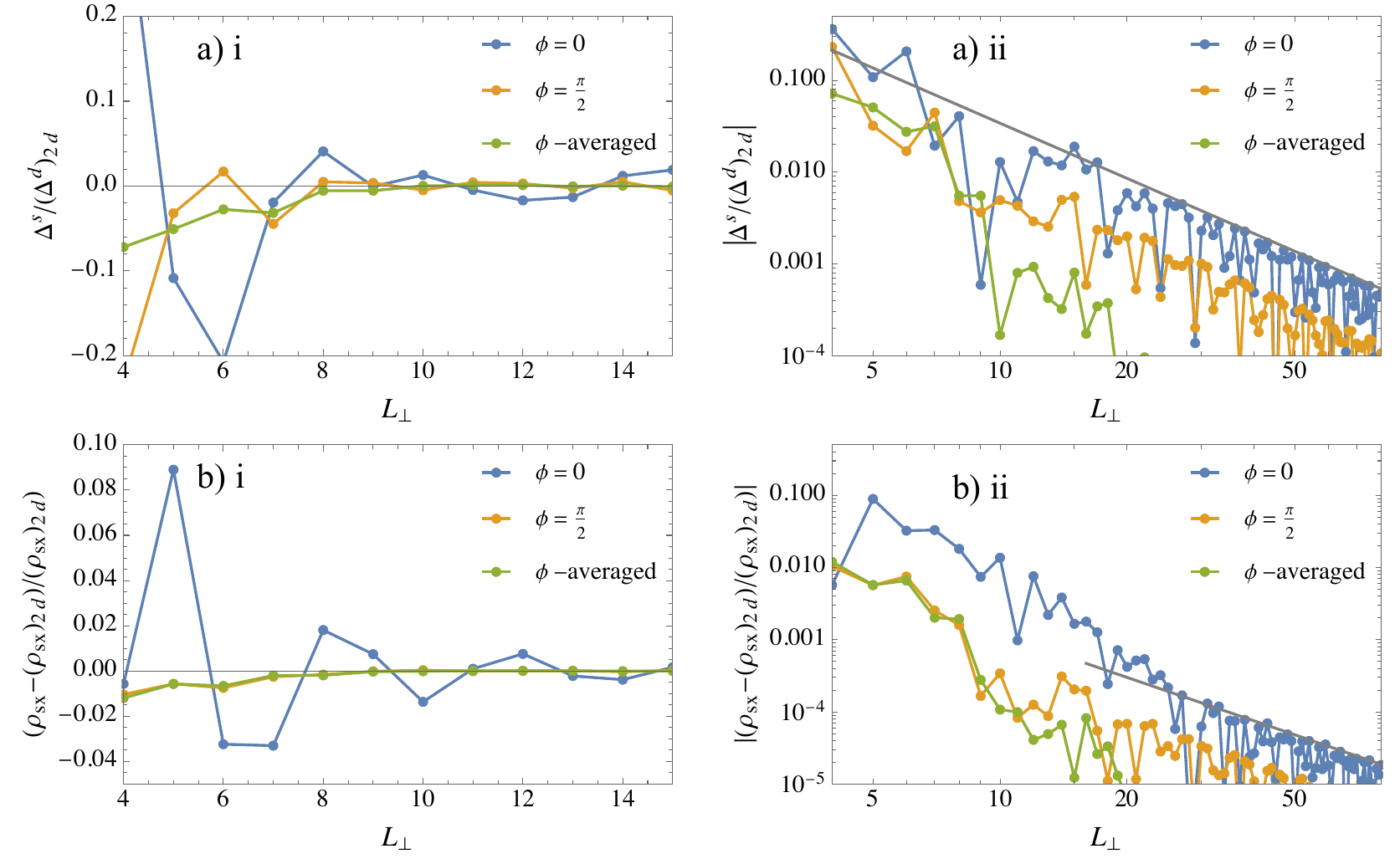}
\caption{
Results of BCS calculations for $\mu = -1.5$, $V_s = -2.5$, and $V_t =- 1.75$. The two-dimensional gap for these parameters takes the form given in Eq.~(\ref{eq:2d_dwave}) with $\Delta_{2d} \approx 0.35$. Panels a)i-ii in the top row depict the $s$-wave gap component normalized by $\Delta_{2d}.$ Panels b)i-ii in the bottom row show the relative error in the superfluid stiffness. a)i and b)i show the ${L_{\perp}}$-dependence of observables at $\phi = 0$ and the special flux value $\phi = \pi/2$, as well as the ${L_{\perp}}$-dependence of $\phi$-averaged observables; a)ii and b)ii show the same thing but on a log-log scale and up to larger ${L_{\perp}}$. The dark gray lines are curves of the form $\text{const}/{L_{\perp}}^{2}$, with the constant fit by hand.
\label{fig:nodal_gap_sf_stiffness}}
\end{figure*}

Fig.(~\ref{fig:nodal_gap_sf_stiffness}) illustrates the ${L_{\perp}}$ and $\phi$-dependence of the gap function and the component of the superfluid stiffness 
parallel to the cylinder\footnote{Note that the superfluid stiffness depends explicitly on the $t$ for $x$-hopping, which is distinct from the $\tau$ for $x$-hopping whenever the interaction is not purely onsite. So although we are holding fixed $\tau$, it is necessary to determine the corresponding $t$ using the appropriate self-consistency equation, which can be found in Appendix~\ref{app:self_consistency}.}. We first describe the top panel, which shows the gap component $\Delta^s$. Needless to say, this quantity vanishes in the 2d limit. We can clearly see $1/L_{\perp}^2$ scaling and the fact that averaging over $\phi$ results in asymptotically faster convergence. We also see 
in panel a)i that simply setting $\phi = \pi/2$ does a relatively good job at speeding up convergence at short distances, although in panel a)ii the advantage of averaging becomes apparent.

We next describe the bottom panel, which shows the finite-size error in the superfluid stiffness. Again we see asymptotic $1/L_{\perp}^2$ scaling, but it sets in only at very large ${L_{\perp}}$.  At smaller circumferences, the finite-size correction decays like it would in a fully-gapped superconductor. In particular,  for small circumferences, setting $\phi = \pi/2$ is 
essentially  equivalent to averaging over twist angles. 

\begin{figure}
\centering
\includegraphics[width=0.8\linewidth]{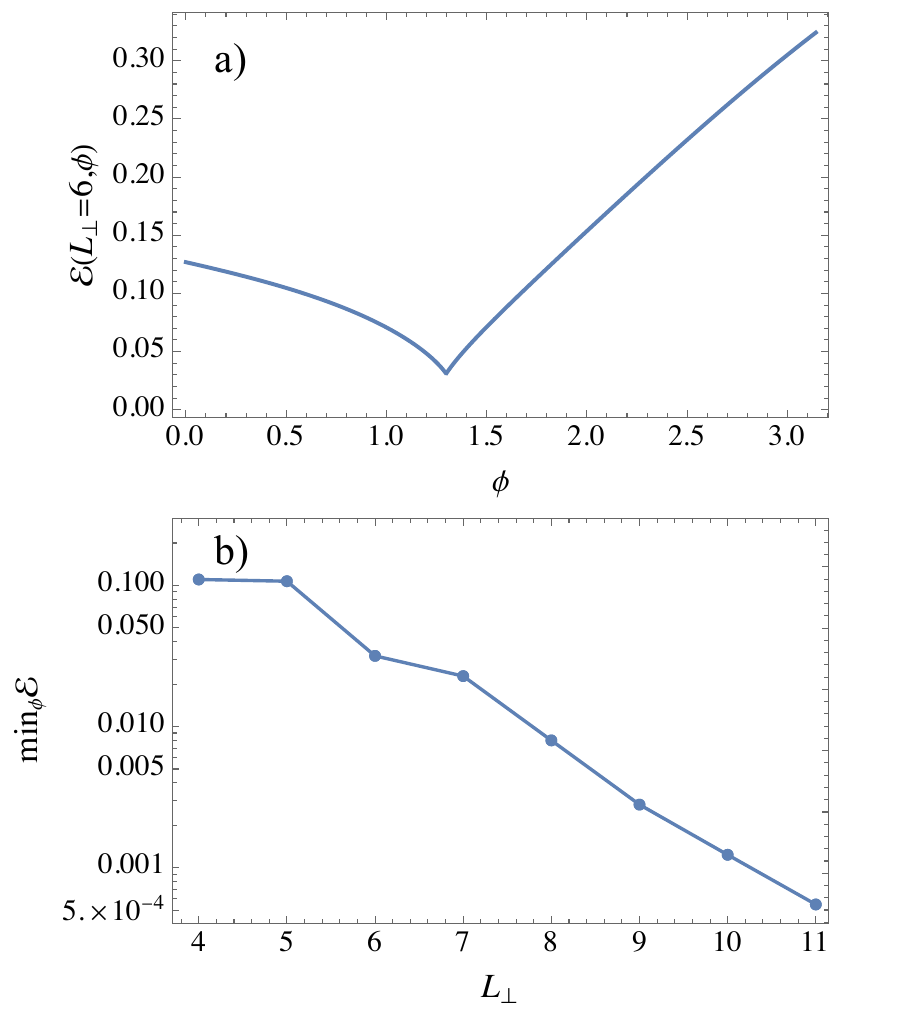}
\caption{
Quasiparticle gap for the same model parameters as Fig.~\ref{fig:nodal_gap_sf_stiffness}. The top panel a) shows $\mathcal{E}({L_{\perp}} = 6, \phi)$ as a function of $\phi$. The bottom panel b) shows $\min_{\phi}\mathcal{E}({L_{\perp}}, \phi)$ versus ${L_{\perp}}$ on a log scale.
\label{fig:nodal_qp_gap}}
\end{figure}

Finally, we turn to the quasiparticle gap. Representative results are depicted in Fig.~\ref{fig:nodal_qp_gap}. We see that at fixed ${L_{\perp}}$, the quasiparticle gap is finite at all $\phi$ but exhibits a deep, cusp-like minimum. (There is, in fact, a small discontinuity in the quasiparticle gap here, but it is well below the scale of the plot.)  We also see that $\min_{\phi} \mathcal{E}({L_{\perp}},\phi)$ vanishes exponentially with a rather short decay length.

\section{Conclusions}
We conclude by summarizing our most important results and presenting some concrete suggestions to DMRG practitioners. 

Firstly, if the 2d limit is any type of superconductor, this should be clear from simply looking at the pair-field correlations at intermediate distances, large compared to ${L_{\perp}}$ but small compared to $\ell_x = A{L_{\perp}} e^{2\pi \rho_s {L_{\perp}}/v}$. Specifically,  as expressed in Eq. \ref{eq:sc_corr}, for  ${L_{\perp}} \ll |x| \ll \ell_x$, the pair-field correlations are essentially equal to those of the 2d system.    This probably does not necessitate keeping the very large bond dimensions needed to properly extract the correct power-law that characterizes the decay of correlations at asymptotically long distances.  

Secondly, twisted boundary conditions are  generally a very useful tool -- see Eqs. \ref{eq:fully_gapped_correction_special_flux},  \ref{eq:fully_gapped_minimum},  \ref{eq:nodal_phi_average}, and Refs~\cite{He_et_al_2017,Hu_et_al_2019}. If one suspects the system is approaching a fully gapped superconductor, then using the special twist angle $\phi = \pm \pi/2$ can substantially reduce finite size errors. More generally, averaging over all $\phi$ can do the same. Twisted boundary conditions are also an excellent way to distinguish between nodal and fully gapped superconductors, as the quasiparticle gap minimized over all $\phi$ in a nodal superconductor will be at worst exponentially small in circumference. From the observation of a 
$\phi$-minimized quasiparticle gap 
 that does not decrease percipitously with increasing ${L_{\perp}}$, one can confidently conclude that the 2d limit is 
 fully gapped -- i.e. does not have nodal quasi-particles.

Finally, anisotropy can substantially speed up the rate at which a fully gapped 
system approaches the 2d limit.  Specifically, if in the 2d limit, the essential nature of the ordered  state (e.g. is it superconducting or not, or whether it is a spin liquid or not) is not sensitive to the specifics of the point-group symmetry one should introduce a sufficient amount of anisotropy to render the underlying Fermi surface  
open.  In this case, the essential correlations of the 2d limit are apparent for  $\xi_{2d} \gg {L_{\perp}}\gg 1$, i.e. even if ${L_{\perp}}$ is small compared to the typical emergent correlation length, $\xi_{2d}$, of the 2d system. 

\begin{acknowledgments}
We thank Prashant Kumar and Michael Zaletel for helpful comments on a draft of this paper. This work was supported by the Department of Energy, Office of Basic Energy Sciences, Division of Materials Sciences and Engineering, under contract DE-AC02-76SF00515. 
\end{acknowledgments}

\appendix 
\begin{widetext}

\section{Superconducting order parameter correlation function}
\label{app:sc_corr}
Since the action~(\ref{eq:sc_action}) is Gaussian, $ \ev{\Phi^*(\vb{r})\Phi(0)} = |\Phi_0|^2 e^{\ev{\theta(\vb{r})\theta(0)}-\langle\theta^2\rangle}$ and
$ \ev{\Phi}_{2d} = \Phi_0 e^{-\langle\theta^2\rangle_{2d}/2}$, so
\begin{equation}
     \ev{\Phi^*(\vb{r})\Phi(0)} = |\ev{\Phi}_{2d}|^2e^{I(\vb{r})},
\end{equation}
where
\begin{equation}
    I(x,y) = \ev{\theta(\vb{r})\theta(0)} - \langle\theta^2\rangle 
+ \langle\theta^2\rangle_{2d}.
\end{equation}
Above, $\vb{r} = (x,y)$, all fields are implicitly being evaluated at the same imaginary time, and expectation values without a subscript are taken at finite ${L_{\perp}}$ whereas those with a 2d subscript are taken at ${L_{\perp}} = \infty$. We also assume, without loss of generality, that $-{L_{\perp}}/2 < y < {L_{\perp}}/2$.

Applying the standard methods for computing such Gaussian expectation values,
\begin{equation}
    I(\vb{r}) = \frac{1}{\rho_s} \bigg[
    \int \frac{ d \omega}{2\pi}
    \int \frac{d k_x}{2\pi}
    \frac{1}{{L_{\perp}}}\sum_{n = 0}^{{L_{\perp}} - 1}
    \left.\frac{(e^{i\vb{k} \vdot \vb{r}} -1)f_c(\omega,\vb{k})}
    {\omega^2/v^2+k^2}\right|_{\vb{k} = (k_x, \frac{2\pi n}{{L_{\perp}}})} +
    \int \frac{ d \omega}{2\pi}
    \int \frac{d^2k}{(2\pi)^2}
    \frac{f_c(\omega,\vb{k})}
    {\omega^2/v^2+k^2}
    \bigg]
    \label{eq:I_def}
\end{equation}
Here, $f_c$ is some function implementing a UV cutoff. Besides the fact that it vanishes appropriately fast as its arguments grow in magnitude, we do not assume any specific form for $f_c$.

Applying the Poisson summation formula to the sum over $n$ in Eq.~(\ref{eq:I_def}) and defining
\begin{equation}
    J(\vb{r}) = \frac{1}{v}\int \frac{ d \omega}{2\pi}
    \int \frac{d^2k}{(2\pi)^2}
    \frac{e^{i \vb{k} \vdot \vb{r}}f_c(\omega,\vb{k})}
    {\omega^2/v^2+k^2},
\end{equation}
we find
\begin{align}
    I(\vb{r}) =  \frac{v}{\rho_s} \bqty{J(\vb{r}) + \sum_{m \ne 0} [J(x,y+ m {L_{\perp}}) - J(0, m {L_{\perp}})] }
\end{align}
Since we are interested in $r \gg 1$, we require the form of $J$ for arguments much larger in magnitude than the cutoff length:
\begin{equation}
    J(\vb{r}) = \frac{1}{4\pi r} + \order{\frac{1}{r^2}}.
\end{equation}
The well-known leading order term is universal (independent of the choice of $f_c$). Thus,
\begin{equation}
    I(\vb{r}) = \frac{v}{4 \pi \rho_s}  \bqty{ \frac{1}{r} -\frac{1}{{L_{\perp}}} R(x/{L_{\perp}}; y/{L_{\perp}}) + -\frac{1}{{L_{\perp}}} R(x/{L_{\perp}}; -y/{L_{\perp}})} + \ldots
    \label{eq:simplified_I}
\end{equation}
where
\begin{equation}
    R(z;\lambda) \equiv \sum_{m>0}\pqty{\frac{1}{{m}} - \frac{1}{\sqrt{(z)^2 + (\lambda +m)^2}}
    },
\end{equation}
and where the terms in Eq.~(\ref{eq:simplified_I}) $\ldots$ fall off like $1/L_{\perp}^2$ or $1/r^2$ or faster.

Consider now the limiting behavior of $R(z;\lambda)$ as $z \to \infty$. Since $R(z;\lambda)$ is roughly a harmonic sum cut off at upper limit $z$, $R(z,\lambda) \sim \log(z) - c + \ldots$ for some constant $c$. It is straightforward to evaluate the constant $c$ numerically. Doing so to an estimated accuracy of 11 decimal places and using an inverse symbolic calculator \cite{Inverse_Symbolic}, we find $c = \log(2) - \gamma$, where $\gamma$ is Euler's constant. Hence, we arrive at the following asymptotic expansion of $I(\vb{r})$ with respect to $x$:
\begin{equation}
    I(\vb{r}) \sim \frac{v}{2\pi \rho_s} \bqty{
    \frac{1}{L_{\perp}} \log(1/\abs{x}) + C(L_{\perp}) + \ldots}  \text{ as } \abs{x} \to \infty,
\end{equation}
where the coefficient $C(L_{\perp})$ satisfies $C(L_{\perp}) = \log(AL_{\perp})/L_{\perp} + \order{L_{\perp}^{-2}}$ and where $A = 2 e^{-\gamma} \approx 1.12$ is universal. This also implies
\begin{equation}
    \ev{\Phi(\vb{r})\Phi(0)} \sim |\ev{\Phi}_{2d}|^2\pqty{\dfrac{A {L_{\perp}}}{\abs{x}}}^{\frac{v}{2\pi \rho_s {L_{\perp}}}},
\end{equation}
as in the main text, where the precise meaning of this relation is
\begin{equation}
    \lim_{\abs{x} \to \infty} \frac{|\ev{\Phi}_{2d}|^2\pqty{\dfrac{A {L_{\perp}}}{\abs{x}}}^{\frac{v}{2\pi \rho_s {L_{\perp}}}}}{\ev{\Phi(\vb{r})\Phi(0)}} = 1 + \order{L_{\perp}^{-2}}.
\end{equation}
\section{BCS theory on a cylinder}
\label{app:general_BCS}
\subsection{The trial Hamiltonian}
\label{app:trial_ham}
To simplify notation, we assign the $x$-direction a finite length $L_{\parallel}$. We introduce the following mode expansion for arbitrary $\vb{r} \in \mathbb{Z}^2$, consistent with the twisted boundary condition~(\ref{eq:identification}):
\begin{equation}
    c_{\vb{r}\sigma} = \frac{1}{\sqrt{L_{\parallel} {L_{\perp}}}} \sum_{n_x=0}^{L_{\parallel} -1}\sum_{n_y=0}^{N_{\perp} -1} e^{i \vb{k} \vdot \vb{r}}c_{\vb{k},\sigma}|_{\vb{k} = (\frac{2\pi}{L_{\parallel}}, \frac{2\pi n +\sigma \phi}{{L_{\perp}}})}.
\end{equation}
Plugging this into the trial Hamiltonian \ref{eq:trial_ham},
\begin{equation}
    H_{\text{trial}} =  \sum_{n_x=0}^{L_{\parallel} -1}\sum_{n_y=0}^{N_{\perp} -1} \left.\bqty{  \sum_{\sigma} \varepsilon(\vb{k})c^{\dg}_{(\sigma\vb{k}),\sigma}c_{(\sigma\vb{k}),\sigma} + \Delta(\vb{k}) c^{\dg}_{\vb{k},\up}c^{\dg}_{-\vb{k},\down} + \Delta^*(\vb{k}) c_{-\vb{k},\down}c_{\vb{k},\up}}\right|_{\vb{k} = (\frac{2\pi}{L_{\parallel}}, \frac{2\pi n +\sigma \phi}{{L_{\perp}}})}.
\end{equation}
This is diagonalized by the usual Bogoliobov transform: we introduce quasiparticle operators $\gamma_{\vb{k}\up}, \gamma_{\vb{k}\down}$ (which have the same restriction on their momenta as electron operators with the same spin polarization) through
\begin{align}
    c_{\vb{k}\up} &= u_{\vb{k}}^{*}\gamma_{\vb{k}\up} + v_{\vb{k}} \gamma^{\dg}_{-\vb{k} \down} \\
    c_{-\vb{k}\down} &= - v_{\vb{k}} \gamma^{\dg}_{\vb{k}\up} + u_{\vb{k}}^{*} \gamma_{-\vb{k}\down}
\end{align}
where 
\begin{align}
    \abs{u_{\vb{k}}}^2 &= \frac{1}{2}
    \pqty{1+\frac{\varepsilon(\vb{k})}{E(\vb{k})}} \\
    \abs{v_{\vb{k}}}^2 &= \frac{1}{2}
    \pqty{1-\frac{\varepsilon(\vb{k})}{E(\vb{k})}}
\end{align}
with $E({\vb{k}}) = \sqrt{\varepsilon(\vb{k}) + |\Delta(\vb{k})|^2}$, with the relative phases such that $u_{\vb{k}}^{*} v_{\vb{k}} = -\Delta(\vb{k})/(2E({\vb{k}}))$. We find the quasiparticle spectrum from the main text,
\begin{equation}
    H_{\text{trial}} =  \sum_{n_x=0}^{N_{\ell} -1}\sum_{n_y=0}^{N_{c} -1} \left.  \sum_{\sigma}
    E(\sigma\vb{k}) \gamma^{\dg}_{\vb{k}\sigma}\gamma_{\vb{k}\sigma}
    \right|_{\vb{k} = (\frac{2\pi}{L_{\parallel}}, \frac{2\pi n +\sigma \phi}{{L_{\perp}}})} + \text{const}
\end{equation}
and we find the basic electron correlators in the ground state of the trial Hamiltonian to be
\begin{align}
    \ev{c^{\dg}_{\vb{k}\sigma} c_{\vb{k}\sigma}}_{\text{trial}}  &=  \frac{1}{2}\pqty{1-\frac{\varepsilon(\sigma{ \vb{k})}}{E(\sigma{ \vb{k}})}}, \\
    \ev{c_{-\vb{k}, \down} c_{\vb{k},\up}}_{\text{trial}} &= -\frac{\Delta(\vb{k})}{2E(\vb{k})}.
\end{align}

The real-space correlators (\ref{eq:F_def}, \ref{eq:G_def}) are then given, in the limit $N_{\parallel} \to \infty$, by
\begin{align}
    F(\vb{r}) = 
    \int \frac{dk_x}{2\pi} \frac{1}{{L_{\perp}}} \sum_{n = 0}^{{L_{\perp}} -1} \left. e^{i \vb{k} \vdot \vb{r}}\frac{(-\Delta(\vb{k}))}{2E_{\vb{k}}}\right|_{\vb{k} = (k_x, \frac{2\pi n + \phi}{{L_{\perp}}})}, \\
    G(\vb{r}) = 
     \int \frac{dk_x}{2\pi} \frac{1}{{L_{\perp}}} \sum_{n = 0}^{{L_{\perp}} -1} \left. e^{i \vb{k} \vdot \vb{r}}\frac{1}{2}\pqty{1-\frac{\varepsilon({\vb{k})}}{E({\vb{k}})}} \right|_{\vb{k} = (k_x, \frac{2\pi n + \phi}{{L_{\perp}}})}.
\end{align}

\subsection{The self-consistency equations}
\label{app:self_consistency}
We derive the self-consistency equations using the finite temperature variational principal; the zero-temperature limit is taken at the end of the derivation. The variational principal says to minimize (with respect to every variational parameter $\chi$ in the trial Hamiltonian) the quantity
\begin{equation}
    F'_{\text{trial}} = F_{\text{trial}} + \langle H - H_{\text{trial}} \rangle_{\text{trial}},
\end{equation}
where $F_{\text{trial}}$ is the free energy of the grand canonical ensemble defined by the trial Hamiltonian \cite{Feynman_1998}. Using the Feynman-Hellman theorem, $\pdv*{F_{\text{trial}}}{\chi} = \ev{\pdv*{H_{\text{trial}}}{\chi}}_{\text{trial}}$, the variational equations reduce to
\begin{equation}
    \frac{1}{\Omega}\pdv{\ev{H}_{\text{trial}}}{\chi} = \sum_{\vb{r}} 2 \tau(\vb{r})\pdv{G^*(\vb{r})}{\chi} + \Delta(\vb{r})\pdv{F^*(\vb{r})}{\chi} +\Delta^*(\vb{r})\pdv{F(\vb{r})}{\chi}.
\end{equation}
where $\Omega$ is the area of the system. It is a matter of tedious though straightforward algebra to show
\begin{equation}
    \frac{\ev{H}_{\text{trial}}}{\Omega}= \sum_{\vb{r}} 2 t(\vb{r}) G^*(\vb{r}) + \sum_{a =\pm} \bqty{ V^a |F^{a}(\vb{r})|^2 + U^a |G^{a}(\vb{r})|^2}
\end{equation}
where
\begin{subequations}
\begin{align}
    F^{\pm}(\vb{r}) &= \frac{1}{2} (F(\vb{r}) + F(-\vb{r})) \\
    G^{\pm}(\vb{r}) &= \frac{1}{2} (G(\vb{r}) + G(-\vb{r}))
\end{align}
\end{subequations}
and
\begin{align}
    &V^+(\vb{r}) = V_s \\ 
    &V^-(\vb{r}) = V_t \\
    &U^+(\vb{r})= 2\delta_{\vb{r},0} \sum_{\vb{r}'} \frac{\pqty{V_s(\vb{r}') + 3 V_t(\vb{r}')}}{4} -\frac{1}{2}\pqty{- V_s(\vb{r}) + 3V_t(\vb{r})} \\
    &U^-(\vb{r}) = - \frac{1}{2}\pqty{V_s(\vb{r}) + V_t(\vb{r}})
\end{align}
Thus the self-consistency equations are
\begin{align}
    &\tilde{\Delta}(\vb{r}) = \sum_{a = \pm }V^a F^a(\vb{r}) \label{eq:gap_eqn}\\
    &\tau(\vb{r}) = t(\vb{r}) + \sum_{a = \pm }U^a G^a(\vb{r}) \label{eq:HF_eqn}.
\end{align}
Eq.~(\ref{eq:gap_eqn}) is Eq.~(\ref{eq:gap_eqn_main}) from the main text in a slightly different notation. Since we choose to hold fixed $\tau$, Eq.~(\ref{eq:HF_eqn}) can mostly be ignored. It is only needed to compute the superfluid stiffness, which depends explicitly upon the $t$ corresponding to our fixed $\tau$.

\subsection{Other bases and vector notation}
It is convenient to allow for a more general basis, i.e. to introduce some orthonormal basis of functions $\tilde{f}_j(\vb{r})$ and expand
\begin{equation}
    \tilde{\Delta}(\vb{r}) = \sum_j \tilde{f}_j(\vb{r}) \Delta_j 
\end{equation}
and 
\begin{equation}
    F(\vb{r}) = \sum_j \tilde{f}_j(\vb{r}) F_j. 
\end{equation}
Here
\begin{subequations}
\label{eq:FG_components}
\begin{align}
    F_j =  
    \int \frac{dk_x}{2\pi} \frac{1}{{L_{\perp}}} \sum_{n = 0}^{{L_{\perp}} -1} \left. f_j^*(\vb{k})\frac{(-\Delta(\vb{k}))}{2E_{\vb{k}}}\right|_{\vb{k} = (k_x, \frac{2\pi n + \phi}{{L_{\perp}}})}, \\
    G_j = \int \frac{dk_x}{2\pi} \frac{1}{{L_{\perp}}} \sum_{n = 0}^{{L_{\perp}} -1} \left. f^*_j(\vb{k})\frac{1}{2}\pqty{1-\frac{\varepsilon({\vb{k})}}{E({\vb{k}})}} \right|_{\vb{k} = (k_x, \frac{2\pi n + \phi}{{L_{\perp}}})},
\end{align}
\end{subequations}
and
\begin{align}
    f_j(\vb{k}) = \sum_{\vb{r}} e^{-i \vb{k} \vdot \vb{r}} \tilde{f}_j(\vb{r}).
\end{align}
In what follows we will also introduce a vector notation, letting e.g. $[\vec{\Delta}]_j = \Delta_j$, and the same whenever else an arrow vector symbol is used. In such a notation the gap equation is 
\begin{equation}
    \vec{\Delta} = \mathsf{V} \vec{F}
\end{equation}
where $\mathsf{V}$ is a Hermitian matrix encoding the interaction. In what follows, we will pick a basis in which each $f_j$ has definite parity and is real. Let us further assume that $\Delta(\vb{k})$ is real, as in the main text. Then $\vec{\Delta}$, $\vec{F}$, $\vec{G}$ and $\mathsf{V}$ are all real, and $\mathsf{V}$ has no matrix elements connecting even and odd components.

\section{Finite-circumference perturbation theory}
\label{app:finite_circumference_perturbation}
Explicitly including all the functional dependencies of $F_j$, the gap equation is 
\begin{equation}
    \vec{\Delta} = \mathsf{V} \vec{F}(\vec{\Delta},{L_{\perp}},\phi). 
\end{equation}
We rearrange this as
\begin{align}
       \vec{\Delta} - \mathsf{V} \vec{F}(\vec{\Delta},\infty,0) &= \mathsf{V}[\vec{F}(\vec{\Delta},{L_{\perp}},\phi) -
       \vec{F}(\vec{\Delta},\infty,0)] 
    \\   & =
       \mathsf{V} \delta' \vec{F}(\vec{\Delta},{L_{\perp}},\phi).
       \label{eq:gap_eqn_rearrange}
\end{align}
Here, we have introduced some notation: for any expectation value $A$ computed in the trial Hamiltonian with the gap fixed to $\vec{\Delta}$,
\begin{equation}
    \delta ' A(\vec{\Delta},{L_{\perp}},\phi^{-1}) \equiv A(\vec{\Delta},{L_{\perp}},\phi) - A(\vec{\Delta},\infty,0).
\end{equation}
We will refer to $\delta ' A$ as the ``explicit shift'' in the quantity $A$.

Next, we define $\delta \vec{\Delta} = \vec{\Delta} - \vec{\Delta}_{2d}$, where $\vec{\Delta}_{2d}$ is the solution to the gap equation in the 2d limit, and expand the left hand side of Eq.~(\ref{eq:gap_eqn_rearrange}) in a power series:
\begin{equation}
     \vec{\Delta} - \mathsf{V} \vec{F}(\vec{\Delta},\infty,0) = (1 + \mathsf{V}\mathsf{M}) \delta \vec{\Delta}  + \order{(\delta \Delta)^2}
     \label{eq:series}
\end{equation}
where
\begin{align}
    \mathsf{M}_{ij} &= -\left.\pdv{F_i(\vec{\Delta},\infty,0)}{\Delta_j}\right|_{\vec{\Delta} = \vec{\Delta}_{2d}} \\
    &= \int \frac{d^2k}{(2\pi)^2} f_i(\vb{k})f_j(\vb{k}) \bqty{  \frac{1}{2\sqrt{\varepsilon^2(\vb{k}) + \Delta_{2d}^2(\vb{k})}} - \frac{\Delta_{2d}^2(\vb{k})}{2(\varepsilon^2(\vb{k}) + \Delta_{2d}^2(\vb{k}))^{3/2}}
    } \\
    &= \int \frac{d^2k}{(2\pi)^2} f_i(\vb{k})f_j(\vb{k}) \frac{\varepsilon^2(\vb{k})}{2(\varepsilon^2(\vb{k}) + \Delta_{2d}^2(\vb{k}))^{3/2}},
\end{align}
so that $\mathsf{M}$ is a positive definite, symmetric matrix. By inverting the series (\ref{eq:series}), we can re-write Eq.(\ref{eq:gap_eqn_rearrange}) as
\begin{equation}
     \delta \vec{\Delta} = \widetilde{\mathsf{V}} \delta' \vec{F}(\vec{\Delta},{L_{\perp}},\phi) + [\text{second and higher powers of } \delta' F_j(\vec{\Delta},{L_{\perp}},\phi)]
     \label{eq:gap_eqn_rearrange2}
\end{equation}
where 
\begin{equation}
    \widetilde{\mathsf{V}} = \frac{1}{1+ \mathsf{V}\mathsf{M}} \mathsf{V}
\end{equation}
Note that $\widetilde{\mathsf{V}}$ is symmetric. This can be made explicit by writing it as
\begin{equation}
    \widetilde{\mathsf{V}} = \mathsf{M}^{-1/2} \bqty{\frac{\mathsf{M}^{1/2}\mathsf{V} \mathsf{M}^{1/2}}{1+\mathsf{M}^{1/2}\mathsf{V} \mathsf{M}^{1/2}}}\mathsf{M}^{-1/2}.
    \label{eq:V_tilde}
\end{equation}
Since the explicit shift vanishes as $L_{\perp} \to \infty$,
the solution to Eq.~(\ref{eq:gap_eqn_rearrange2}) has the expansion
\begin{equation}
     \delta \vec{\Delta} = \widetilde{\mathsf{V}} \delta' \vec{F}(\vec{\Delta}_{2d},{L_{\perp}},\phi) + \ldots
     \label{eq:gap_eqn_lowest_order}
\end{equation}
where the terms in $\ldots$ are subleading. In what follows we will typically truncate Eq.~(\ref{eq:gap_eqn_lowest_order}) at the leading order term.

The series implicit in Eq.~(\ref{eq:gap_eqn_lowest_order}) presumes that as $L_{\perp} \to \infty$, there is exactly one solution which approaches $\vec{\Delta}_{2d}$. In fact, for a nodal superconductor there can be multiple solutions that approach $\vec{\Delta}_{2d}$ as $L_{\perp} \to \infty$. However, we will see in Appendix~\ref{app:qp_gap} that the splitting between solutions is negligibly small. Thus the lowest-order truncation of Eq.~(\ref{eq:gap_eqn_lowest_order}) works well for most purposes.

\subsection{Invertibility of $(1+ \mathsf{M}^{1/2}\mathsf{V}\mathsf{M}^{1/2})$}
The perturbation theory described above is valid only if $(1+\mathsf{M}^{1/2}\mathsf{V}\mathsf{M}^{1/2})$ is invertible. To see that it is indeed invertible, consider the Hessian matrix of the energy of the 2d state. We find that\footnote{To be explicit, the Hessian matrix is comprised of four blocks: the diagonal blocks $(1/\Omega)\pdv*{\ev{H}_{\text{trial}}}{\Delta_j}{\Delta_k}$, $(1/\Omega)\pdv*{\ev{H}_{\text{trial}}}{\tau_j}{\tau_k}$, and the off-diagonal block $(1/\Omega)\pdv*{\ev{H}_{\text{trial}}}{\Delta_j}{\tau_k}$ and its transpose. The latter is of order $ \Delta \log(\Delta)$ (as compared to the diagonal blocks which are of order $\log(\Delta)$ and $1$, respectively) so we treat the Hessian as block-diagonal. Moreover, in  Eq.~(\ref{eq:Hessian}) we have omitted sub-leading terms that are of order $ \Delta^2 \log(\Delta)$.
}
\begin{equation}
    \left.\frac{1}{\Omega}\pdv{\ev{H}_{\text{trial},2d}}{\Delta_j}{\Delta_k}\right|_{\vec{\Delta} = \vec{\Delta}_{2d}} =
    2 [ \mathsf{M}+ \mathsf{M}\mathsf{V} \mathsf{M}]_{jk} = 
    2 [ \mathsf{M}^{1/2}(1+\mathsf{M}^{1/2}\mathsf{V} \mathsf{M}^{1/2})\mathsf{M}^{1/2}]_{jk}.
    \label{eq:Hessian}
\end{equation}
Since we are perturbing about the ground state of the 2d limit, the Hessian and thus $(1+ \mathsf{M}^{1/2}\mathsf{V}\mathsf{M}^{1/2})$ must be positive definite. In other words, $(1+ \mathsf{M}^{1/2}\mathsf{V}\mathsf{M}^{1/2})$ has all positive eigenvalues and is invertible.

\subsection{Corrections to other observables}
The finite-size correction to any expectation value $A$ in the trial Hamiltonian can be expanded as
\begin{equation}
    A(\vec{\Delta},{L_{\perp}},\phi) - A_{2d} = \sum_j\left.\pdv{A(\vec{\Delta},\infty,0)}{\Delta_j}\right|_{\vec{\Delta} = \vec{\Delta}_{2d}} \delta \Delta_{j} + \delta' A(\vec{\Delta}_{2d},{L_{\perp}},\phi) + \ldots
    \label{eq:general_correction}
\end{equation}
By Wick's theorem, the leading order contribution to $\delta' A(\vec{\Delta}_{2d},{L_{\perp}},\phi)$ is a linear combination of $\delta'F_j(\vec{\Delta}_{2d},{L_{\perp}},\phi)$ and $\delta'G_j(\vec{\Delta}_{2d},{L_{\perp}},\phi)$. Additionally, we showed above that $\delta \vec{\Delta} = \mathsf{V} \delta' \vec{F}(\vec{\Delta}_{2d},{L_{\perp}},\phi)$. It follows that the leading-order finite-size correction to any expectation value $A$ is a linear combination of the explicit shifts $\delta'F_j(\vec{\Delta}_{2d},{L_{\perp}},\phi)$ and $\delta'G_j(\vec{\Delta}_{2d},{L_{\perp}},\phi)$. In what follows, we assume that $\varepsilon(\vb{k})$ is an even function and that $\vec{\Delta}_{2d}$ contains only even components. Then the finite-size corrections to quantities that are even (odd) under inversion are linear combinations of  $\delta'F_j(\vec{\Delta}_{2d},{L_{\perp}},\phi)$ and $\delta'G_j(\vec{\Delta}_{2d},{L_{\perp}},\phi)$ such that $f_j$ is even (odd).

\subsection{Evaluating the explicit corrections to $F$ and $G$.}
Applying the Poisson summation formula to Eq.~(\ref{eq:FG_components}) we arrive at 
\begin{subequations}
\label{eq:explicit_corrections}
\begin{align}
    \delta' F_j(\vec{\Delta},{L_{\perp}},\phi)
    &= \sum_{m \ne 0}e^{-i m \phi} \mathcal{F}_j(\vec{\Delta},m{L_{\perp}}), \\
     \delta' G_j(\vec{\Delta},{L_{\perp}},\phi)
    &= \sum_{m \ne 0} e^{-i m \phi} \mathcal{G}_j(\vec{\Delta},m{L_{\perp}}),
\end{align}
\end{subequations}
where
\begin{subequations}
\label{eq:script_FG}
\begin{align}
    \mathcal{F}_j(\vec{\Delta},y)
    &= \int \frac{d^2 k}{(2\pi)^2} e^{i y k_y} f_j(\vb{k}) \frac{(-\Delta(\vb{k}))}{2E(\vb{k})}, \\
     \mathcal{G}_j(\vec{\Delta},y)
    &= \int \frac{d^2 k}{(2\pi)^2} e^{i y k_y} f_j(\vb{k}) \frac{1}{2}\pqty{1-\frac{\varepsilon({\vb{k})}}{E({\vb{k}})}}.
\end{align}
\end{subequations}
Note that each of $\mathcal{F}_j(\vec{\Delta}_{2d},y )$ and $\mathcal{G}_j(\vec{\Delta}_{2d},y )$
is a 2d correlation functions at separation $y \vu{y}$, convolved with the basis function $\tilde{f}_j$.  Eqs.~(\ref{eq:explicit_corrections}) and (\ref{eq:script_FG}) thus give the precise relation between finite-circumference corrections -- which are proportional to the explicit shifts in $F$ and $G$ -- and 2d correlation functions along the circumferential direction. Note that $\mathcal{F}_j(\vec{\Delta}_{2d},y)$ or $\mathcal{G}_j(\vec{\Delta}_{2d},y)$ is an even (odd) function of $y$ when $f_j$ is even (odd).

\section{The fully gapped case}
\label{app:fully_gapped}
In this case each of $\mathcal{F}_j$ and $\mathcal{G}_j$ are Fourier transforms of functions analytic at real momenta. As a result, they decay exponentially:
\begin{equation}
    \mathcal{F}_j(\vec{\Delta}_{2d}, y) \sim \mathcal{G}_j(\vec{\Delta}_{2d}, y) \sim
    e^{-\abs{y}/{\xi_{\perp}}}
    \begin{cases}
    \cos[{Q_{\perp}} y + \sgn(y)\alpha] & \text{even component} \\
    \sin[{Q_{\perp}} y + \sgn(y)\alpha] & \text{odd component} 
    \end{cases}
\end{equation}
Note that while ${Q_{\perp}}$ and ${\xi_{\perp}}$ are the same for all $\mathcal{F}_j$ and $\mathcal{G}_j$ (they are properties of the zero structure of $E_{2d}(\vb{k})$ in the complex plane -- see below), the phase shift $\alpha$ is understood to be unique to the quantity under consideration.

Plugging this form into (\ref{eq:explicit_corrections}), the exponential decay implies that terms with $|m| >1$ are subleading. The exception is for even components at $\phi = \pm \pi/2$, in which case the $|m| = 1$ terms cancel and the $|m|=2$ terms lead. Explicitly,
\begin{equation}
    \delta' F_j(\vec{\Delta}_{2d},{L_{\perp}},\phi) \sim
    \delta' G_j(\vec{\Delta}_{2d},{L_{\perp}},\phi) \sim
    e^{-{L_{\perp}}/{\xi_{\perp}}}
    \begin{cases}
    \cos(\phi)e^{-{L_{\perp}}/{\xi_{\perp}}}\cos({Q_{\perp}} {L_{\perp}} + \alpha) & \text{even component} \\ 
    \sin(\phi)e^{-{L_{\perp}}/{\xi_{\perp}}}\sin({Q_{\perp}} {L_{\perp}} + \alpha) & \text{odd component} \\ 
    \end{cases}
\end{equation}
and for even components at $\phi = \pm \pi/2$, 
\begin{equation}
     \delta' F_j(\vec{\Delta}_{2d},{L_{\perp}},\pm \pi/2) \sim
    \delta' G_j(\vec{\Delta}_{2d},{L_{\perp}},\pm \pi/2) \sim e^{-2{L_{\perp}}/{\xi_{\perp}}}\cos(2{Q_{\perp}} {L_{\perp}} + \alpha).
\end{equation}
Since the finite size correction to any expectation value $A$ is a linear combination of $\delta ' F_j$ and $\delta ' G_j$ of the appropriate spatial parity, this implies Eq.~(\ref{eq:fully_gapped_correction}) and Eq.~(\ref{eq:fully_gapped_correction_special_flux}).

\subsection{The correlation length $\xi_{\perp}$}
\label{app:xi_y}

\label{app:xi_perp}
Here we show that in the small gap limit, $\xi_{\perp}$ behaves qualitatively differently depending upon whether or not there is a point on the Fermi surface at which $\vb{v}_F$ points along $\vu{y}$. Let us determine $\xi_{\perp}$ by evaluating $\mathcal{F}_j$.
\begin{equation}
    \mathcal{F}_j(\vec{\Delta}_{2d},y) = \int \frac{d k_y}{2\pi} e^{i k_y y} \widetilde{\mathcal{F}}_j(\vec{\Delta}_{2d}, k_y),
\end{equation}
where
\begin{equation}
    \widetilde{\mathcal{F}}_j(\vec{\Delta}_{2d}, k_y) = - \int \frac{d k_x}{2\pi} \frac{f_j(\vb{k}) \Delta_{2d}(\vb{k})}{2 E_{2d}(\vb{k})}
    \label{eq:sc_correlator_aux}
\end{equation}
The Fourier transform $\widetilde{\mathcal{F}}_j(\vec{\Delta}_{2d}, k_y)$ will be non-analytic at various points in the complex $k_y$-plane. If $k'_y$ is the point of non-analyticity with the smallest positive imaginary part, then $k'_y = Q_{\perp} + i/\xi_{\perp}$. 

The function $\widetilde{\mathcal{F}}_j(\vec{\Delta}_{2d}, k_y)$ is an integral with respect to $k_x$ of a function of $(k_x,k_y)$ that is analytic everywhere except at isolated singular points -- the zeros of the quasiparticle dispersion $E_{2d}(\vb{k})$. (We assume an analytic dispersion and gap function.) It is known that in this situation, $\widetilde{\mathcal{F}}_j(\vec{\Delta}_{2d}, k_y)$ is non-analytic at a point $k_y'$ if and only if two solutions $k_x(k_y)$ to $E_{2d}(\vb{k}) = 0$ pinch the $k_x$-integration contour as $k_y \to k_y'$ \cite{Eden_et_al_2002}. A necessary condition for such a pinching to occur at $k_y$ is for $(k_x, k_y)$ to solve $E_{2d}(\vb{k}) = 0$ as well as $\partial_{k_x}[\varepsilon^2(\vb{k})+ \Delta^2(\vb{k})] = 0$. Assuming for simplicity a constant gap function, these equations reduce to
\begin{subequations}
\label{eq:pinch}
\begin{align}
    &\varepsilon(\vb{k}) = \pm i \Delta_{2d}, \\
    &\partial_{k_x} \varepsilon(\vb{k}) = 0.
\end{align}
\end{subequations}
If $\vb{v}_F$ points along $\vu{y}$ at some momentum on the (real) Fermi surface, then the solution to Eq.~(\ref{eq:pinch}) has an imaginary part proportional to $\Delta_{2d}$, leading to $\xi_{\perp} \propto 1/\Delta_{2d}$. On the other hand, if $\vb{v}_F$ never points along $\vu{y}$ on the (real) Fermi surface then the solution to Eq.~(\ref{eq:pinch}) has an imaginary part that is finite even at $\Delta_{2d} = 0$ (i.e. even in a Fermi liquid state), leading to $\xi_{\perp}$ on the order of a lattice constant for any $\Delta_{2d}$.

For the dispersion $\varepsilon(\vb{k}) = - 2 t_x\cos(k_x) -2t_y\cos(k_y) - \mu$ appearing in Eq.~(\ref{eq:simple_model}), the relevant solution is 
\begin{equation}
    (k_x',k_y') = (0, \arccos((- i \Delta_{2d} - \mu - 2 t_x)/2t_y)),
\end{equation}
yielding ${\xi_{\perp}} = 1/\Im[\arccos((- i \Delta_{2d} - \mu - 2 t_x)/2t_y)]$.

\section{The nodal case}
\label{app:nodal}
In this case, $\mathcal{F}_j$ and $\mathcal{G}_j$ are Fourier transforms of functions non-analytic at the nodal points, resulting in a power law decay. The large-$|y|$ behavior receives a contribution from each node. Let $\vb{v}_F(\vb{k}) = \nabla_{\vb{k}} \varepsilon(\vb{k})$ and $\vb{v}_{\Delta}(\vb{k})  = \nabla_{\vb{k}} \Delta(\vb{k})$ denote the Fermi and gap function velocities, respectively. Then
\begin{subequations}
\label{eq:nodal_FG}
\begin{align}
    \mathcal{F}_j(\vec{\Delta},y) &\sim
    \sum_{\text{nodes } \vb{Q}=  (Q_{\parallel},Q_{\perp})} f_j(\vb{Q})e^{iQ_{\perp} y} \int \frac{d ^2 q}{(2\pi)^2} \frac{e^{i q_y y}(- \vb{v}_{\Delta}(\vb{Q}) \vdot \vb{q})}
    {2\sqrt{
    (\vb{v}_F(\vb{Q}) \vdot \vb{q})^2
    +(\vb{v}_{\Delta}(\vb{Q}) \vdot \vb{q})^2
    }} \nonumber \\
    &=-i\sum_{\text{nodes } \vb{Q} = (Q_{\parallel},Q_{\perp})} \frac{f_j(\vb{Q})\det[\vb{v}_F(\vb{Q}), \vb{v}_{\Delta}(\vb{Q})]v_{F,x}(\vb{Q})}{ (v_{Fy}^2(\vb{Q})+v_{\Delta y}^2(\vb{Q}))^{3/2}} \frac{\sgn{y}}{4 \pi y^2}e^{iQ_{\perp} y}
    \\
    \mathcal{G}_j(\vec{\Delta},y) &\sim
    \sum_{\text{nodes } \vb{Q}=  (Q_{\parallel},Q_{\perp})} f_j(\vb{Q})e^{iQ_{\perp} y} \int \frac{d ^2 q}{(2\pi)^2} \frac{e^{i q_y y}(- \vb{v}_F(\vb{Q}) \vdot \vb{q})}
    {2\sqrt{
    (\vb{v}_F(\vb{Q}) \vdot \vb{q})^2
    +(\vb{v}_{\Delta}(\vb{Q}) \vdot \vb{q})^2
    }} \nonumber \\
    &=-i\sum_{\text{nodes } \vb{Q} = (Q_{\parallel},Q_{\perp})}
     \frac{f_j(\vb{Q})\det[\vb{v}_{\Delta}(\vb{Q}), \vb{v}_{F}(\vb{Q})]v_{\Delta,x}(\vb{Q})}{ (v_{Fy}^2(\vb{Q})+v_{\Delta y}^2(\vb{Q}))^{3/2}} \frac{\sgn{y}}{4 \pi y^2}e^{iQ_{\perp} y} 
\end{align}
\end{subequations}
where $\det(\vb{v}_1,\vb{v}_2) \equiv v_{1x}v_{2y} - v_{1y}v_{2x}$. The nodal momenta are implicitly functions of $\vec{\Delta}$. Plugging Eq.~(\ref{eq:nodal_FG}) into Eq.~(\ref{eq:explicit_corrections}), this leads to
\begin{subequations}
\label{eq:nodal_explicit_corrections}
\begin{align}
    \delta 'F_j(\vec{\Delta}, {L_{\perp}},\phi) = \sum_{\text{nodes } \vb{Q} = (Q_{\parallel},Q_{\perp})} \frac{f_j(\vb{Q})\det[\vb{v}_F(\vb{Q}), \vb{v}_{\Delta}(\vb{Q})]v_{F,x}(\vb{Q})}{ (v_{Fy}^2(\vb{Q})+v_{\Delta y}^2(\vb{Q}))^{3/2}} \frac{1}{2 \pi L_{\perp}^2}\text{Cl}_2({L_{\perp}} Q_{\perp}-\phi) \\
\delta 'G_j(\vec{\Delta}, {L_{\perp}},\phi) = \sum_{\text{nodes } \vb{Q} = (Q_{\parallel},Q_{\perp})}
 \frac{f_j(\vb{Q})\det[\vb{v}_{\Delta}(\vb{Q}), \vb{v}_{F}(\vb{Q})]v_{\Delta,x}(\vb{Q})}{ (v_{Fy}^2(\vb{Q})+v_{\Delta y}^2(\vb{Q}))^{3/2}} \frac{1}{2 \pi L_{\perp}^2}\text{Cl}_2({L_{\perp}} Q_{\perp}-\phi) 
\end{align}
\end{subequations}
where the Clausen function of order 2 is defined in Eq.~(\ref{eq:clausen_def}).

Under the assumption that the nodes in the 2d limit are $\pm (Q_{\parallel,2d}, Q_{\perp,2d})$ and $\pm (-Q_{\parallel,2d}, Q_{\perp,2d})$, we find
\begin{subequations}
\begin{align}
    \delta 'F_j(\vec{\Delta}_{2d}, {L_{\perp}},\phi) =  \frac{f_j(\vb{Q}_{2d})\det[\vb{v}_F(\vb{Q}_{2d}), \vb{v}_{\Delta}(\vb{Q}_{2d})]v_{F,x}(\vb{Q}_{2d})}{ (v_{Fy}^2(\vb{Q}_{2d})+v_{\Delta y}^2(\vb{Q}_{2d}))^{3/2}} \frac{1}{\pi L_{\perp}^2}[\text{Cl}_2({L_{\perp}} Q_{\perp,2d}-\phi) \pm \text{Cl}_2({L_{\perp}} Q_{\perp,2d}+\phi)] \\
\delta 'G_j(\vec{\Delta}_{2d}, {L_{\perp}},\phi) = 
 \frac{f_j(\vb{Q}_{2d})\det[\vb{v}_{\Delta}(\vb{Q}_{2d}), \vb{v}_{F}(\vb{Q}_{2d})]v_{\Delta,x}(\vb{Q}_{2d})}{ (v_{Fy}^2(\vb{Q}_{2d})+v_{\Delta y}^2(\vb{Q}_{2d}))^{3/2}} \frac{1}{\pi L_{\perp}^2}[\text{Cl}_2({L_{\perp}} Q_{\perp,2d}-\phi) \pm \text{Cl}_2({L_{\perp}} Q_{\perp,2d}+\phi)]
\end{align}
\end{subequations}
Here the upper (lower) sign is for even (odd) components. This leads to Eq.~(\ref{eq:nodal_expect_correction}) from the main text.

\subsection{The quasiparticle gap}
\label{app:qp_gap}
As described in the main text, the case in which the quasiparticle gap is finite for all $\phi$ corresponds to a first-order transition, i.e. a situation in which there are multiple solutions to the gap equation which cross in energy at some critical $\phi$. As discussed above, the possibility of multiple solutions approaching $\vec{\Delta}_{2d}$ in the limit $L_{\perp} \to \infty$ is something which is missed by Eq.~(\ref{eq:gap_eqn_lowest_order}). 

In this section, we address the splitting of the solutions to the gap equation by truncating the right hand side of Eq.~(\ref{eq:gap_eqn_rearrange2}) at first order and solving the resulting non-trivial equation for $\vec{\Delta}$. On a cylinder, the nodal momenta can be parameterized as $\pm \vb{Q}^{\pm} = \pm (Q_{\parallel}^{\pm},Q_{\perp}^{\pm})$ and $\pm (-Q_{\parallel}^{\pm},Q_{\perp}^{\pm})$, such that $\vb{Q}^{\pm} \to \vb{Q}_{2d}$ in the 2d limit. With this in mind, we work with the following approximation to Eq.~(\ref{eq:gap_eqn_rearrange2}):
\begin{equation}
    \delta \vec{\Delta}=   \frac{\det[\vb{v}_F(\vb{Q}_{2d}), \vb{v}_{\Delta}(\vb{Q}_{2d})]v_{F,x}(\vb{Q}_{2d})}{ (v_{Fy}^2(\vb{Q}_{2d})+v_{\Delta y}^2(\vb{Q}_{2d}))^{3/2}} \frac{1}{\pi L_{\perp}^2}[\widetilde{\mathsf{V}}\vec{f}(\vb{Q}_{2d})\text{Cl}_2({L_{\perp}} Q_{\perp}^{+}-\phi) +\widetilde{\mathsf{V}}\vec{f}'(\vb{Q}_{2d}) \text{Cl}_2({L_{\perp}} Q_{\perp}^{-}+\phi)].
    \label{eq:nodal_approx}
\end{equation}
where $f_j'$ = $f_j$ if $f_j$ is even and $f_j' = -f_j$ if $f_j$ is odd. On the right hand side above, the nodal momenta are evaluated in the 2d limit everywhere except inside the Clausen functions. This is legitimate because only the Clausen functions have a non-analytic dependence on nodal momenta.

Note that $\vec{\Delta}$ enters the right hand side of Eq.~(\ref{eq:nodal_approx}) only through $Q^{\pm}_{\perp}$. Thus, to lowest order we can reduce Eq.~(\ref{eq:nodal_approx}) to a pair of equations for $Q_{\perp}^{\pm}$ by taking the dot product 
of both sides with $\pdv*{Q_{\perp}^\pm}{\vec{\Delta}}$. It is straightforward to show that
\begin{align}
    \pdv{Q_{\perp}^+}{\vec{\Delta}} &= \frac{- \vec{f}(\vb{Q}_{2d}) v_{Fx}(\vb{Q}_{2d})}{\det[\vb{v}_F(\vb{Q}_{2d}), \vb{v}_{\Delta}(\vb{Q}_{2d})]}, \\
    \pdv{Q_{\perp}^-}{\vec{\Delta}} &= \frac{-\vec{f}'(\vb{Q}_{2d}) v_{Fx}(\vb{Q}_{2d})}{\det[\vb{v}_F(\vb{Q}_{2d}), \vb{v}_{\Delta}(\vb{Q}_{2d})]}.
\end{align}
Taking the dot product of this with (\ref{eq:nodal_approx}), we find the  coupled equations
\begin{equation}
    Q_{\perp}^{\pm} - Q_{\perp,2d} = \frac{1}{L_{\perp}^2} \bqty{\beta \text{Cl}_2({L_{\perp}} Q_{\perp}^\pm \mp \phi) + \beta' \text{Cl}_2({L_{\perp}} Q_{\perp}^\mp \pm \phi)}.
    \label{eq:nodal_self_consistency}
\end{equation}
where
\begin{equation}
    \beta = \frac{-v_{Fx}^2(\vb{Q}_{2d})}{\pi(v_{Fy}^2(\vb{Q}_{2d})+v_{\Delta y}^2(\vb{Q}_{2d}))^{3/2}} \vec{f}(\vb{Q}_{2d})^T \widetilde{\mathsf{V}}\vec{f}(\vb{Q}_{2d}),
    \label{eq:nodal_beta}
\end{equation}
and there is a similar expression for $\beta'$.

Without loss of generality, let us consider the upper node $Q^{+}_{\perp}$. There is always some $\phi^{\star}$ such that there is a solution to Eq.~(\ref{eq:nodal_self_consistency}) satisfying ${L_{\perp}} Q_{\perp}^+ = \phi^{\star} + 2\pi n$ for integer $n$, i.e. such that the upper node intersects some allowed momentum slice. Without loss of generality, assume $n = 0$. We now address the question of whether or not there exist additional lower energy solutions that do not intersect the allowed momentum slice.

Let us introduce $D^{+} \equiv {L_{\perp}} Q^{+} - \phi$, which is $L_{\perp}$ times the distance along the $y$-axis from the upper node to the nearest allowed momentum slice. When $D^{+}$ is small, the quasiparticle gap is proportional to $|D^{+}|/L_{\perp}$. The Clausen function of order 2 has a logarithmically divergent slope at argument $ 2 n \pi$:
\begin{equation}
    \text{Cl}_2(2n\pi + \theta) \sim \theta\log(e/\abs{\theta}).
    \label{eq:clausen_asymptotic}
\end{equation}
Using this, for $\phi$ in the vicinity of $\phi^{\star}$ and small $D^{+}$, we can reduce the pair of equations in Eq.~(\ref{eq:nodal_self_consistency}) to the following equation for $D^+$ alone:
\begin{equation}
    D^{+} + (\phi - \phi^{*})\bqty{1+ \order{\frac{1}{L_{\perp}}}}
    = \frac{\beta}{{L_{\perp}}} D^{+}\log(e/|D^{+}|) + \order{\frac{1}{L_{\perp}^2}}
    \label{eq:nodal_equation_reduced}
\end{equation}

For large $L_{\perp}$ there are two cases:
\begin{itemize}
    \item $\beta < 0$: As a function of $\phi$ there is a single solution branch such that $D^+$ vanishes at $\phi = \phi^{\star}$. In this case $\min_{\phi} \mathcal{E}(L_{\perp},\phi) = 0$.
    
    \item $\beta > 0$: As a function of $\phi$ there are three solution branches in the vicinity of $\phi = \phi^{\star}$. See Fig.~\ref{fig:nodal_solution_schematic}. On the middle (green) branch, $D^+$ vanishes at $\phi = \phi^{\star}$. But this branch is unstable. (This must be the case because it can ``annihilate'' with the upper (blue) and lower (orange) branches, each of which is stable since it is the only solution for $\phi$ far from $\phi^{\star}$.) Instead, it is always the case that one of the upper (blue) and lower (orange) branches has lowest energy. $D^+$ does not change sign on either the upper or lower branch, but for $\phi$ such that these two branches coexist, $\abs{D^+}= \order{e^{-{L_{\perp}}/\beta}}$ on either one. In this case $\min_{\phi} \mathcal{E}(L_{\perp},\phi) = \order{e^{-{L_{\perp}}/\beta}}$. We see that $\ell_n$ from the main text is equal to $\beta$.
\end{itemize}
Note that by Eq.~(\ref{eq:V_tilde}), the matrix $\widetilde{\mathsf{V}}$ is negative definite whenever $\mathsf{V}$ is negative definite -- a condition we referred to as ``fully attractive'' in the main text. Thus, Eq.~(\ref{eq:nodal_beta}) implies that for a fully attractive interaction, $\beta > 0$, and thus the quasiparticle gap is finite for all $\phi$. (But its minimum value is exponentially small in $L_{\perp}$.)

\begin{figure}
\centering
\includegraphics[width=0.5\linewidth]{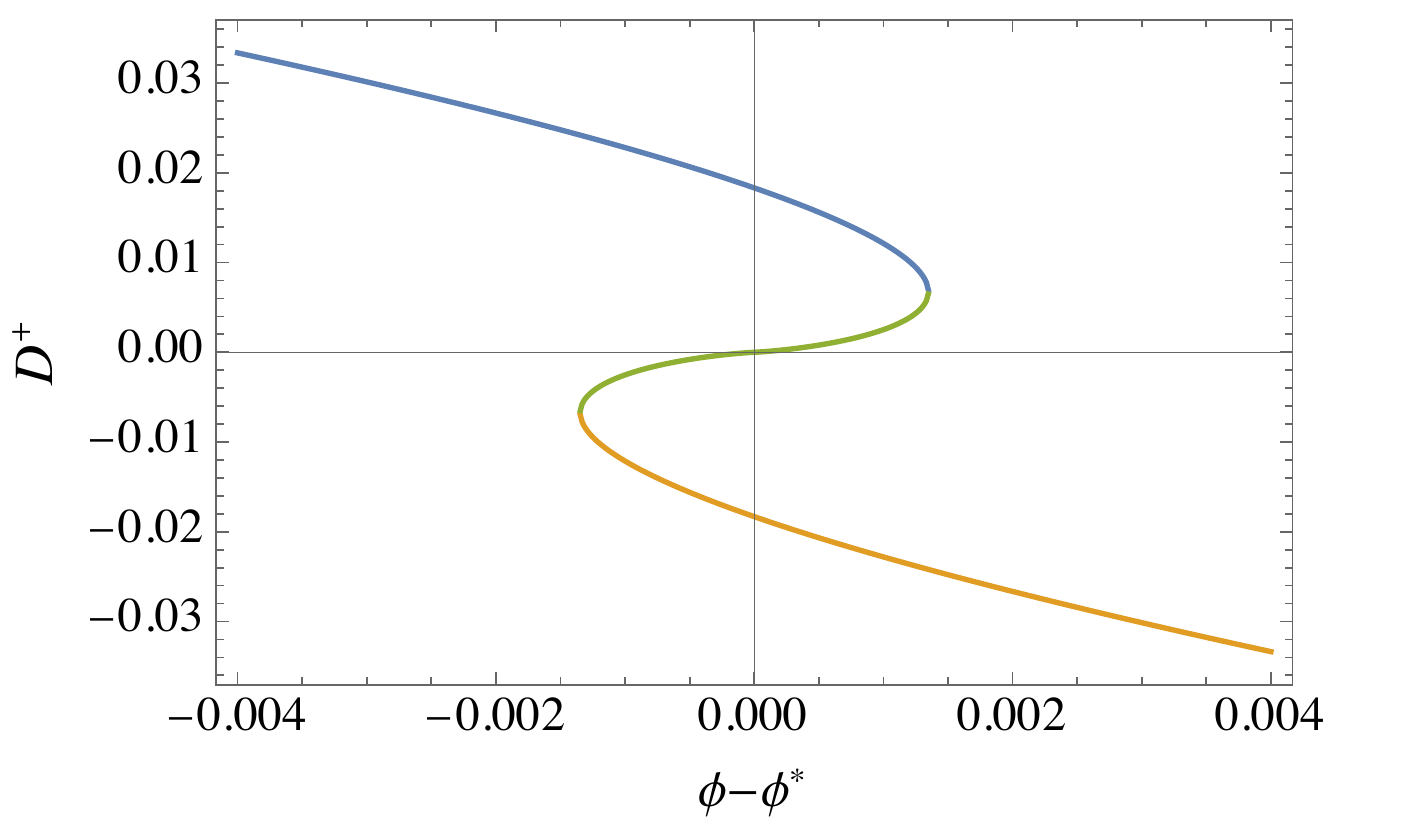}
\caption{
Solutions to Eq.~(\ref{eq:nodal_equation_reduced}) for $\beta/L_{\perp} = 0.2$. The solutions are only schematic, in the sense that terms inside the $\mathcal{O}$ symbol in Eq.~(\ref{eq:nodal_equation_reduced}) are neglected.
\label{fig:nodal_solution_schematic}}
\end{figure}
\end{widetext}

\bibliography{bibliography}

\end{document}